\documentclass[a4paper,fleqn]{cas-sc}

\usepackage[numbers]{natbib}

\usepackage{subcaption}
\usepackage{graphicx}
\usepackage{amsmath}
\usepackage{amsthm}
\usepackage{amssymb}
\usepackage{multirow}
\usepackage{url}

\begin{document}
\let\WriteBookmarks\relax
\def\floatpagepagefraction{1}
\def\textpagefraction{.001}

\title [mode = title]{Reputation assimilation mechanism for sustaining cooperation}

\shorttitle{Reputation assimilation for sustaining cooperation}
\shortauthors{S. He et~al.}

\author[1]{Siyu~He}
\affiliation[1]{organization={College of Artificial Intelligence, Southwest University},
                addressline={}, 
                city={Chongqing},
                postcode={400715}, 
                country={PR China}}
\credit{Writing – review \& editing, Writing – original draft, Methodology, Investigation, Formal analysis
}

\author[2]{Qin~Li}[
                        orcid=0000-0001-5480-0992
                        ]
\cormark[1]
\ead{qinli1022@swu.edu.cn}
\credit{Writing – review \& editing, Validation, Methodology, Formal analysis, Conceptualization}
\affiliation[2]{organization={Business College, Southwest University},
                city={Chongqing},
                postcode={402460}, 
                country={PR China}}

\author[1]{Minyu~Feng}
\credit{Writing – review \& editing, Writing – original draft, Supervision, Methodology, Funding acquisition}

\author[3]{Attila~Szolnoki}
\credit{Writing – review \& editing, Funding acquisition, Formal analysis, Conceptualization}
\affiliation[3]{organization={Institute of Technical Physics and Materials Science, Centre for Energy Research},
                addressline={P.O. Box 49}, 
                city={Budapest},
                postcode={H-1525}, 
                country={Hungary}}

\cortext[cor1]{Corresponding author}

\begin{abstract}
Keeping a high reputation, by contributing to common efforts, plays a key role in explaining the evolution of collective cooperation among unrelated agents in a complex society. Nevertheless, it is not necessarily an individual feature, but may also reflect the general state of a local community. Consequently, a person with a high reputation becomes attractive not just because we can expect cooperative acts with higher probability, but also because such a person is involved in a more efficient group venture. These observations highlight the cumulative and socially transmissible nature of reputation. Interestingly, these aspects were completely ignored by previous works. To reveal the possible consequences, we introduce a spatial public goods game in which we use an assimilated reputation simultaneously characterizing the individual and its neighbors' reputation. Furthermore, a reputation-dependent synergy factor is used to capture the high (or low) quality of a specific group. Through extensive numerical simulations, we investigate how cooperation and extended reputation co-evolve, thereby revealing the dynamic influence of the assimilated reputation mechanism on the emergence and persistence of cooperation.
By fostering social learning from high-reputation individuals and granting payoff advantages to high-reputation groups via an adaptive multiplier, the assimilated reputation mechanism promotes cooperation, ultimately to the systemic level.
\end{abstract}


\maketitle

\section{Introduction}
Cooperation and defection are two fundamental modes of interaction that can be frequently detected in both nature and human societies~\cite{nowak2006five,zeng2025complex,zhang2025spatial}. From the perspective of short-term self-interest, defectors can gain a higher payoff by free-riding on the contributions of cooperators, which appears consistent with Darwinian selection for individually optimal strategies~\cite{zhu2024reputation,nowak1992evolutionary}. Nevertheless, numerous studies have shown that structural constraints or temporal limitations can fundamentally alter the outcomes of social dilemmas, allowing cooperation to persist even under adverse conditions~\cite{zhang2024limitation}. Empirical evidence further reveals that cooperation remains essential for the survival and stability of biological populations and human communities~\cite{zeng2025bursty,rand2013human,wei2025indirect}. Therefore, understanding the mechanisms that allow cooperation to emerge and persist in the face of individual incentives for defection has long been a central question in different disciplines, including biology, sociology, physics, and game theory~\cite{feng2024information,han_ai22}.

To uncover the intrinsic patterns governing the emergence of cooperation under social dilemmas characterized by conflicts between individual rationality and collective interests, evolutionary game theory has gradually become one of the key research frameworks~\cite{ling2025supervised,han_jrsif22,pi2025dynamic}. In particular, the spatial public goods game (SPGG) has become a canonical model for studying collective cooperation~\cite{perc2013evolutionary,chen2014probabilistic,chen2016individual,han_j_c24}. In traditional SPGG, individuals are located on structured networks. Cooperators contribute to the public pool at a personal cost, while defectors refuse to contribute but share the resulting benefits. From the perspective of individual optimization, choosing defection enables individuals to evade the costs associated with cooperation while securing a ``free-riding'' advantage, a characteristic that makes defection the rationally optimal short-term choice. However, when all individuals prioritize the pursuit of local optima, public resources tend to deplete gradually due to insufficient maintenance efforts. Ultimately, this depletion leads the entire system to fall into the ``tragedy of the commons''~\cite{roca2009evolutionary}.
To mitigate the dilemma, previous studies have introduced various mechanisms such as reward and punishment systems~\cite{fowler2005altruistic,hauert2007via,szolnoki2010reward}, voluntary participation~\cite{hauert_s02}, and social exclusion~\cite{liu_lj_srep17,szolnoki_pre17}. These mechanisms typically draw on theoretical foundations, such as network reciprocity or group selection, to enhance cooperative tendencies. Nevertheless, a critical limitation of most existing mechanisms is that they rely heavily on external rules to either constrain or incentivize individual behavior. Furthermore, the dynamic characterization of feedback derived from social evaluation, which is an essential aspect of practical social interactions, is not adequately addressed in current research~\cite{han_jrsif15,szolnoki2009topology}.

In practical human interactions, reputation acts as a social evaluation metric that reflects an individual's long-term behavioral patterns and plays a critically important role in making decisions and shaping social interactions~\cite{milinski2002reputation,fujimoto2023evolutionary,nowak1998evolution,yue_h_csf25}. Extensive studies have revealed that reputation and reciprocity are deeply intertwined, together forming the cornerstone of cooperative evolution in both biological and social systems~\cite{xia2023reputation}. Recent advances in behavioral game theory further highlight the essential role of communication and shared information in shaping cooperative norms, even in the era of artificial intelligence~\cite{capraro2024language}. Individuals with a high reputation are more likely to gain the trust of others, and their behavioral strategies are also more likely to be copied by their peers~\cite{szolnoki2013correlation,manrique2021psychological}. Nevertheless, the reputation mechanisms incorporated in existing SPGG models still suffer from several limitations. Many studies evaluate behavior using only immediate actions, ignoring how reputation accumulates over repeated interactions~\cite{panchanathan2004indirect,ohtsuki2004should,xie_k_csf25}. In contrast, reputation is frequently tied to the resources an individual holds. For instance, in team cooperation, high-reputation individuals are more likely to access premium resources, leading to greater payoffs. It is also more likely that they are surrounded by akin partners; therefore, they can form a powerful and effective joint venture. Although certain existing models integrate the impact of reputation on strategy learning processes, they still fail to capture the critical positive feedback inherent in real social systems. In this feedback, a high reputation first facilitates an individual’s acquisition of higher payoff, and then the increased payoff further reinforces the individual’s tendency to engage in cooperative behaviors~\cite{chen2015first,szolnoki2015conformity}. These limitations hinder a realistic representation of reputation formation mechanisms and constrain a deeper understanding of cooperative evolution under social dilemmas.

To address the issues mentioned above, we propose a model in which an assimilated reputation is considered, which simultaneously characterizes the actual reputation of the focal player and its close neighborhood. This extended reputation directly affects an individual's propensity, which neighbors imitate by increasing the likelihood that they adopt strategies from higher-reputation neighbors. On the other hand, this reputation dynamically modulates the multiplication factor of the actual public goods game, thereby influencing the collective income of the group. That is to say, reputation updates are based on both an individual's own cooperative behavior and the overall reputation level of neighbors. Individuals tend to form and adjust their social evaluations through interactions with those around them. Besides, members of high-reputation groups are more likely to maintain and accumulate positive reputations, whereas those in low-reputation groups are more susceptible to negative evaluations. Through numerical simulations, a systematic investigation is conducted to examine how dynamics driven by the extended reputation affect the evolution of cooperation under various parameter settings, thus revealing the critical role of reputation in sustaining and propagating the cooperative behavior.

The remainder of our paper is organized as follows. Section~\ref{model} details the game rules, reputation update, and the strategy learning mechanisms of the model. Section~\ref{results} presents the simulation results and analyzes how various parameters influence the fraction of cooperators and the average reputation of the system. Finally, Section~\ref{Conclusions} summarizes the main findings of the model and discusses its theoretical implications.

\section{Model of evolutionary dynamics} \label{model}
In our societies, the willingness of individuals to cooperate is shaped by both immediate payoffs and social evaluations, such as reputation. For example, in team collaborations, individuals with consistently high reputations are more likely to gain trust and support from others, and their behaviors are more readily imitated. To model this phenomenon, we propose an SPGG model that incorporates an assimilated reputation mechanism to characterize how reputation jointly influences payoff amplification and cooperative propensity. Reputation evolves dynamically based on both personal behavior and the average reputation of neighbors, creating a link between past actions and the surrounding social environment.

\subsection{The spatial public goods game}
In the SPGG model, there are $N$ individuals. Initially, each node is assigned the cooperator or defector strategy with equal probability, and its reputation value is uniformly distributed within the interval $[R_{min}, R_{max}]$. During a single round of the public goods game (PGG), each node organizes one PGG group centered on itself and simultaneously participates in the groups organized by its neighbors. The total contribution of each group is multiplied by a factor that depends on the reputation of the central node. Specifically, the multiplication factor $r^{g}{(t)}$ of the group $g$ centered on node $i$ at the time step $t$ is defined as
\begin{equation} \label{eq:r}
    r^{g}{(t)} = r_0 + \beta \left[ \frac{R_i(t)}{R_{max}-R_{min}} \right],
\end{equation}
where $r_0$ denotes the basic synergy factor and $R_i(t)$ represents the reputation of the central node $i$ in the group $g$ at time step $t$. $R_{\rm max}$ and $R_{\rm min}$ are defined as the upper and lower bounds of reputation, respectively. The parameter $\beta$ characterizes the improvement of the synergy factor due to reputation. In particular, a larger value of $\beta$ leads to a stronger effect of the central node's reputation on the amplification of group benefit. In practical terms, this means that when the central node has a high reputation, the same total contributions made by the group can generate a higher shared benefit for all members within the group. In this way, we apply a group-specific synergy factor\cite{lee_hw_csf23}.

In each PGG group, the focal node and its neighbors simultaneously decide whether to contribute to the public pool. Cooperators (C) contribute $c=1$ to the public pool, while defectors (D) contribute nothing. The amplified total contribution is then equally distributed among all members of the group. Therefore, after a single round of the public goods game, the total payoff of node $i$ can be expressed as
\begin{equation}
\Pi_{i}=\sum_{g \in G_{i}}\left(\frac{r^{g} \cdot C_{g} \cdot c}{k_{i}+1}-c \cdot s_{i}\right),
\end{equation}
where $G_i$ denotes the set of PGG groups in which node $i$ participates, $C_g$ represents the number of cooperators in group $g$, $r^{g}$ represents the payoff synergy factor of group $g$, $k_i$ is the degree of central node $i$, and $s_i$ indicates the strategy adopted by node $i$, with $s_i = 1$ for cooperation and $s_i = 0$ for defection.

\subsection{Assimilated reputation mechanism}
In our model, reputation is introduced as a core metric to quantify the social recognition of cooperative behavior, dynamically reflecting a long-term tendency of a node toward strategy selection. Unlike mechanisms relying solely on immediate actions or static evaluations, we introduce a dynamic framework for reputation assessment that incorporates both historical accumulation and feedback from the real-time environment. The framework incorporates the reputation accumulated from past behavior, as well as the influence of the current social environment. The latter is reflected by the average reputation of neighboring nodes. The reputation derived from historical behavior reflects the impact of previously accumulated reputation, while the component based on neighbors’ average reputation reflects how individuals are shaped by the prevailing atmosphere of their surrounding group. In this way, we can provide a more realistic representation of reputation formation in real-world social systems. The update rule for the reputation of individual $i$ can be expressed as
\begin{equation} \label{eq:reputation}
R_{i}(t)=\left\{\begin{array}{ll}
\alpha R_{i}(t-1)+(1-\alpha) \overline{R_{J}(t-1)}+\delta & \text { if } \mathrm{s}_{i}(t)=C \\
\alpha R_{i}(t-1)+(1-\alpha) \overline{R_{J}(t-1)}-\delta & \text { if } \mathrm{s}_{i}(t)=D
\end{array},\right.
\end{equation}
where $\overline{R_{J}(t-1)}$ denotes the average reputation of individual $i$’s neighbors at time step $t-1$, reflecting the local reputational environment surrounding node $i$, with $J$ indexing the neighbor set.
The parameter $\alpha \in [0,1]$ is related to the effect of reputation assimilation. To be specific, when $\alpha = 1$, the player’s reputation is entirely determined by its own previous reputation. In contrast, when $\alpha = 0$, the player’s reputation is predominantly shaped by the average reputation of its neighbors, a characteristic that reflects the social influence analogous to the proverb ``birds of a feather flock together''. For the intermediate range of $0 < \alpha < 1$, the reputation update is jointly determined by the individual’s cooperative behavior and the average reputation of its neighbors. The concept of locally averaging a player-specific quantity, such as payoff~\cite{szolnoki_njp21}, can also be interpreted as a limited perception about other competitors. Furthermore, the parameter $\delta$ represents the reputation perturbation factor, which determines the magnitude of reputation adjustments following individual behaviors. A larger $\delta$ implies a more dramatic change in reputation, leading to greater differentiation between cooperators and defectors during the evolutionary process. The mechanism further motivates nodes to maintain a high reputation through cooperation, as doing so helps them secure long-term payoff advantages. To prevent reputation values from exceeding the initial range $[R_{min}, R_{max}]$, the following truncation rule is applied:
\begin{equation}
    R_{i}(t)=\left\{\begin{array}{cl}
0, & \alpha R_{i}(t-1)+(1-\alpha) \overline{R_{J}(t-1)} \pm \delta<R_{min} \\
\alpha R_{i}(t-1)+(1-\alpha) \overline{R_{J}(t-1)} \pm \delta, & R_{min} \leq \alpha R_{i}(t-1)+(1-\alpha) \overline{R_{J}(t-1)} \pm \delta \leq R_{max} \\
100, & \alpha R_{i}(t-1)+(1-\alpha) \overline{R_{J}(t-1)} \pm \delta>R_{max}
\end{array}\right. . 
\end{equation}

\subsection{Strategy updating}
In the traditional strategy imitation process, after participating in the public goods game, node $i$ randomly selects one of its neighbors (denoted as individual $j$) to imitate its strategy. However, in reality, the strategy learning process is not entirely random, as individuals tend to consider the reputational characteristics of their neighbors when selecting a learning target. Consequently, neighbors with a high reputation are more likely to be imitated due to the higher social recognition of their behavior~\cite{zhang2025evolution,feng2024evolutionary}. Therefore, in our model, the probability of selecting a neighbor $j$ is defined as
\begin{equation} \label{eq:P_j}
    P_{j}=\frac{Z_{(n)}}{Z_{(m)}},
\end{equation}
where $Z_{(n)}=\exp\left(\lambda \cdot \frac{R_j - R_i}{R_{\max}-R_{\min}}\right)$ represents individual $i$’s selection tendency toward neighbor $j$ driven by the reputation difference and $Z_{(m)}=\sum_{k=1}^{G-1} \exp\left(\lambda \cdot \frac{R_k - R_i}{R_{\max}-R_{\min}}\right)$ aggregates the selection tendencies of all neighbors and normalizes them to ensure that the probabilities sum up to 1. Here, $\lambda$ is the reputation sensitivity coefficient. When $\lambda = 0$, node $i$ selects all neighbors with equal probability. As $\lambda$ increases, the focal player is more likely to choose a neighbor with a higher reputation as a learning target. This implies that high-reputation individuals are more likely to be imitated, forming a bias in strategy evolution that is driven by reputation.

The probability that node $i$ adopts the strategy of the selected neighbor $j$ is defined by using the Fermi function:
\begin{equation}
    W_{s_i \rightarrow s_j} = \frac{1}{1 + e^{-\kappa (\Pi_j - \Pi_i)}},
\end{equation}
where $\Pi_j - \Pi_i$ denotes the payoff difference between node $i$ and its neighbor, and $\kappa $ represents a noise level. Accordingly, the imitation becomes deterministic in the limit $\kappa \to 0$ and random in the limit $\kappa \to \infty$.  

\begin{figure}
    \centering
    \includegraphics[width=1\linewidth]{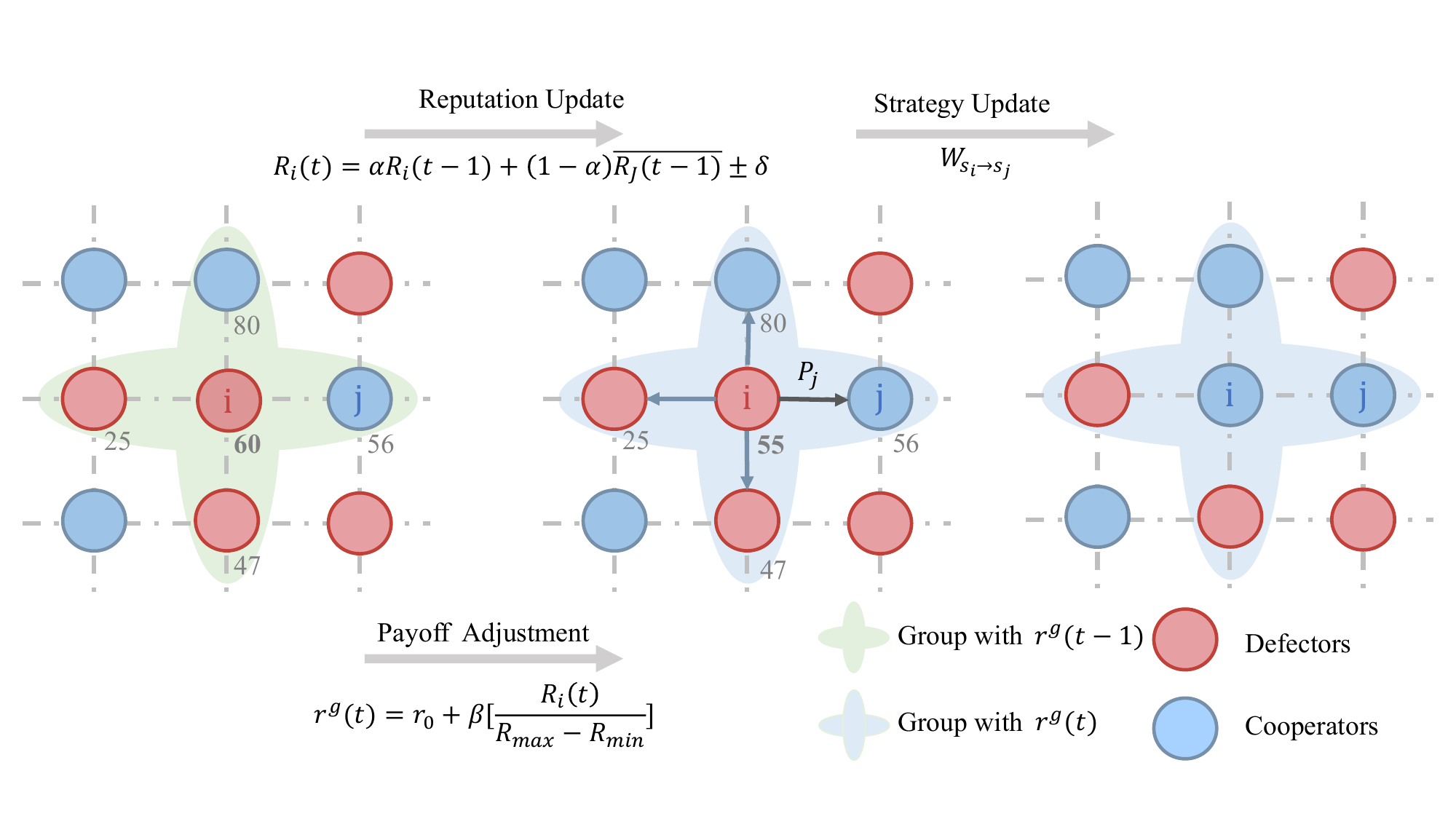}
    \caption{\textbf{Strategy, reputation update, and payoff adjustment process.} Blue nodes represent cooperators, while red nodes represent defectors. The shading indicates different game states: green shading corresponds to the game state at time step $t-1$, and blue shading represents the updated game state. The gray numbers next to the nodes indicate each node's reputation. The model consists of three processes:  \textbf{(a) Reputation update:} The central node updates its reputation based on its current behavior.   \textbf{(b) Payoff adjustment:} After the central node's reputation is updated, the multiplication factor of the PGG group is adjusted according to the updated reputation of the central node.  \textbf{(c) Strategy update:} The central node $i$ selects a neighbor $j$ with probability $P_j$, and adopts the strategy of node $j$ with probability $W_{s_i \rightarrow s_j}$. }
    \label{fig:model}
\end{figure}

The dynamical rules defined above are composed of three main components. Reputation Update, Payoff Adjustment, and Strategy Update, as illustrated in Fig.~\ref{fig:model}. We take the square lattice as an example to illustrate the sequential evolution process in which each player undergoes reputation updating, payoff adjustment, and strategy update within its local neighborhood. For instance, when player $i$ adopts the defection strategy, its reputation decreases according to the update rule given in Eq.~(\ref{eq:reputation}), falling from 60 at time step $t-1$ to 55 at time step $t$. After updating the reputation of player $i$, the multiplication factor of the corresponding PGG group is adjusted in accordance with Eq.~(\ref{eq:r}) based on the central player’s updated reputation. Finally, player $i$ selects a neighbor $j$ following Eq.~(\ref{eq:P_j}) and adopts $j$’s strategy with probability $W_{s_i \rightarrow s_j}$, eventually becoming a cooperator. Notably, when $\beta = \delta = \lambda = 0$ and $\alpha = 1$, the model reduces to the traditional SPGG.

\section{Results and discussions} \label{results}
In this section, we conduct Monte Carlo (MC) simulations to numerically investigate the evolution of cooperation under the proposed rules. The simulation is implemented on a regular square lattice of size $L\times L=50\times 50$ with periodic boundary conditions and a von Neumann neighborhood. Each run lasts for $3 \times 10^3$ Monte Carlo Steps (MCS), and the results are averaged over the final $2 \times 10^3$ MCS to ensure statistical stability. To enhance robustness, each simulation is repeated ten times with different initial conditions, and the reported results represent the averages over these independent realizations. The noise level is fixed at $\kappa = 2$. Each individual's initial reputation values are uniformly distributed within the interval $[0,100]$.

\subsection{Assimilated reputation mechanism across dilemma intensities}
\begin{figure}
    \centering
    \begin{subfigure}[b]{0.48\textwidth}
        \centering
        \includegraphics[width=\linewidth]{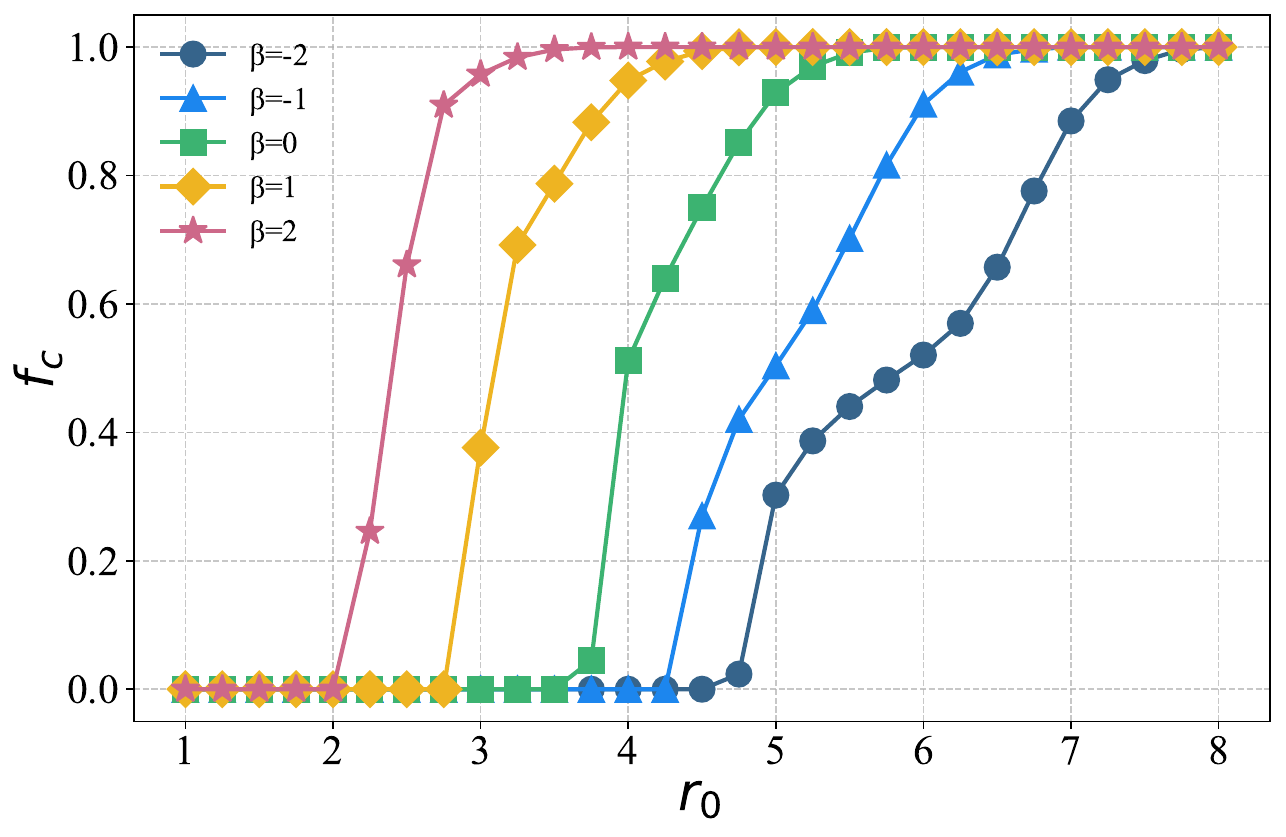}
        \caption{$\lambda=0$}
        \label{fig:lamuda_0_fc}
    \end{subfigure}
    \hfill
    \begin{subfigure}[b]{0.48\textwidth}
        \centering
        \includegraphics[width=\linewidth]{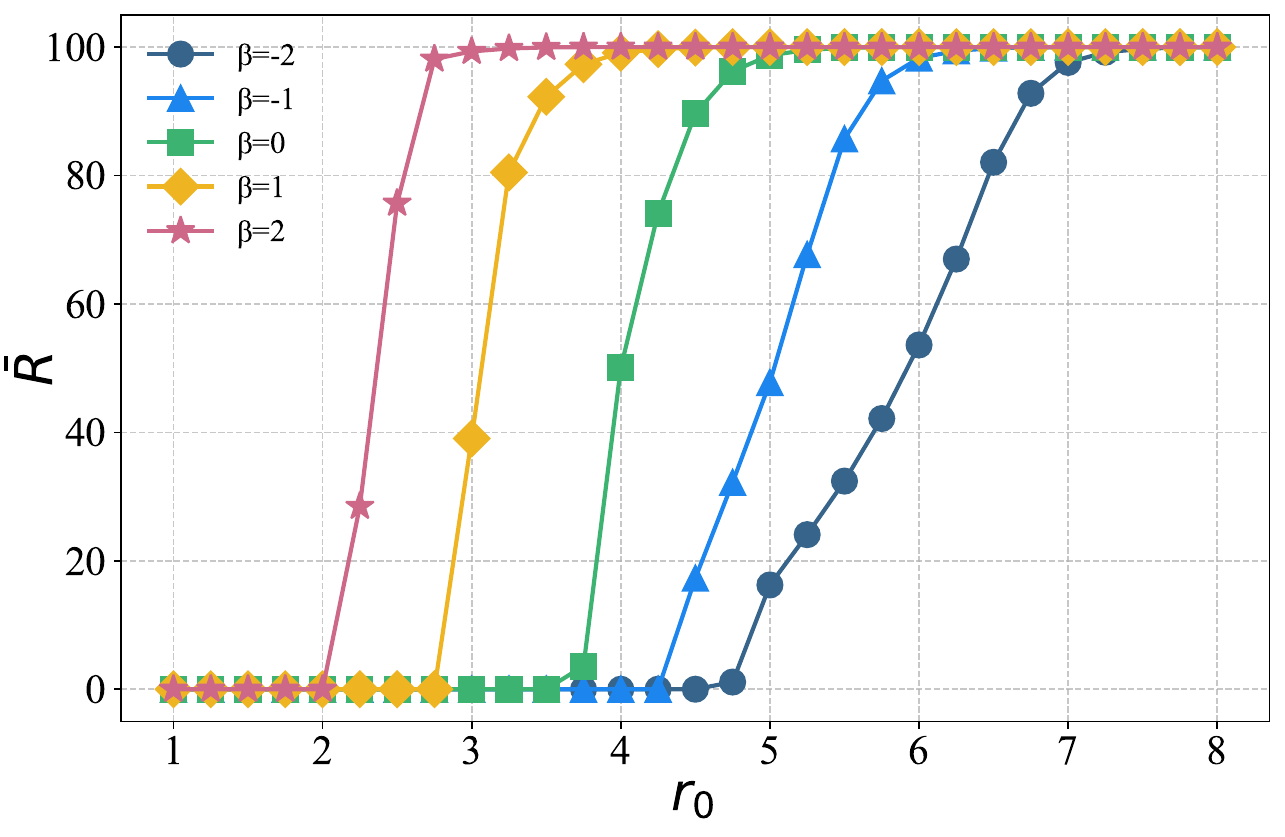} 
        \caption{$\lambda=0$}
        \label{fig:lamuda_0_R}
    \end{subfigure}
    
    \vspace{1em} 

    \begin{subfigure}[b]{0.48\textwidth}
        \centering
        \includegraphics[width=\linewidth]{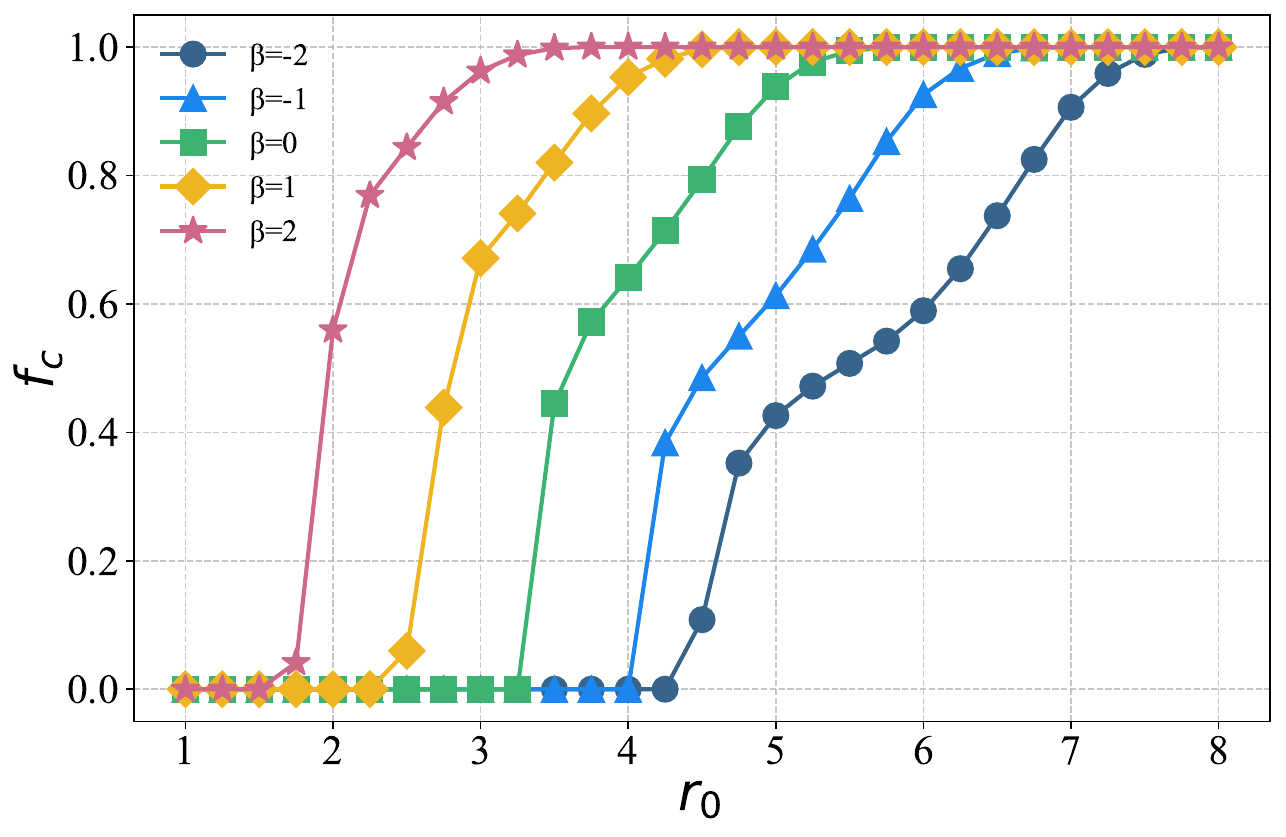}
        \caption{$\lambda=2$}
        \label{fig:lamuda_2_fc}
    \end{subfigure}
    \hfill
    \begin{subfigure}[b]{0.48\textwidth}
        \centering
        \includegraphics[width=\linewidth]{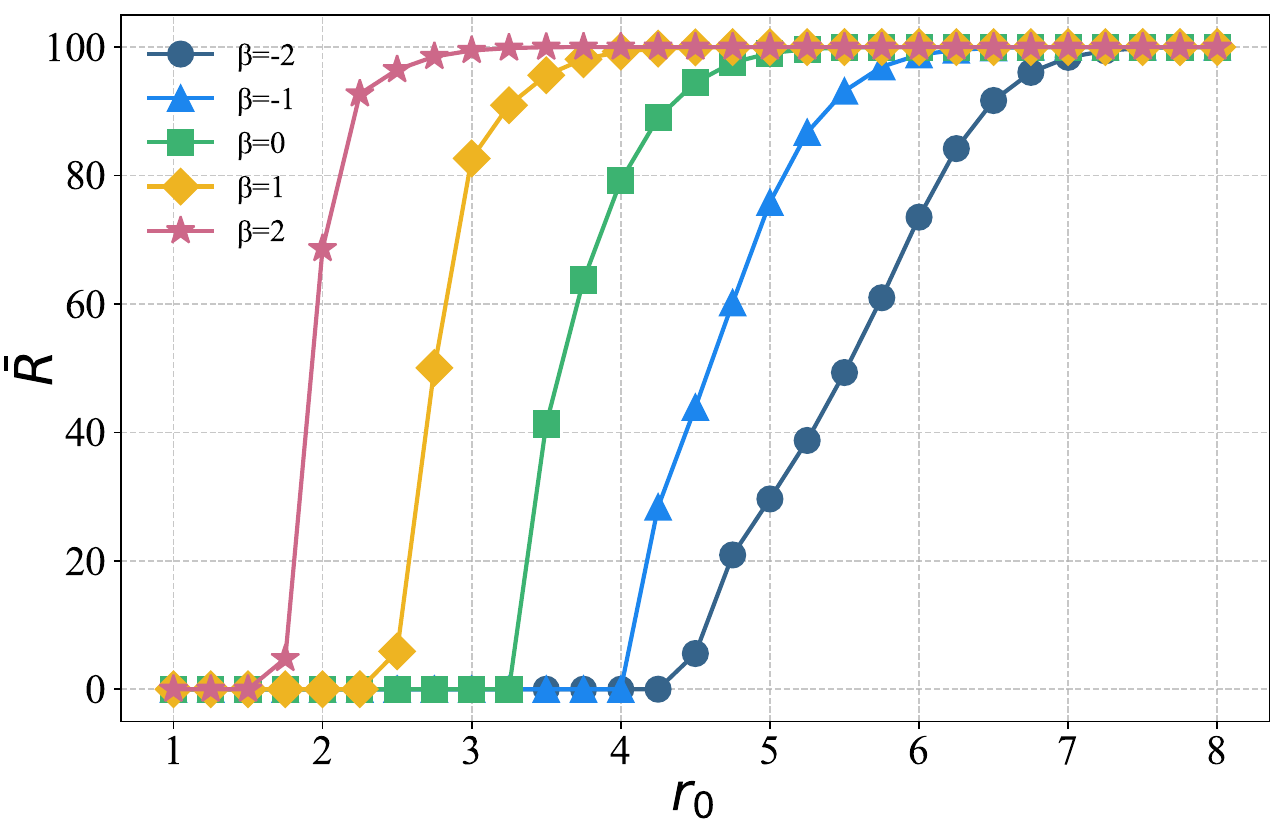} 
        \caption{$\lambda=2$}
        \label{fig:lamuda_2_R}
    \end{subfigure}
    \caption{\textbf{Dependence of $f_c$ and $\bar{R}$ on the $r_0$ under different $\lambda$ and $\beta$ values.} Other parameters are set as $\alpha=0.5$ and $\delta=5$. In Figs.~\ref{fig:lamuda}(\subref{fig:lamuda_0_fc}) and \ref{fig:lamuda}(\subref{fig:lamuda_0_R}), $\lambda=0$, while in Figs.~\ref{fig:lamuda}(\subref{fig:lamuda_2_fc}) and \ref{fig:lamuda}(\subref{fig:lamuda_2_R}), $\lambda=2$. Figs.~\ref{fig:lamuda}(\subref{fig:lamuda_0_fc}) and \ref{fig:lamuda}(\subref{fig:lamuda_2_fc}) show the fraction of cooperators $f_c$ as a function of $r_0$ under different payoff enhancement parameters $\beta$, whereas Figs.~\ref{fig:lamuda}(\subref{fig:lamuda_0_R}) and \ref{fig:lamuda}(\subref{fig:lamuda_2_R}) show the corresponding average reputation $\bar{R}$. Different colored lines represent different $\beta$ values as shown in the legend.}
    \label{fig:lamuda}
\end{figure}

To evaluate the effectiveness of the proposed assimilated reputation mechanism in promoting cooperation compared to other reputation mechanisms across different levels of social dilemma intensity, the influence of these mechanisms on the evolution of cooperation and reputation is first examined under varying baseline synergy factor, as shown in Fig.~\ref{fig:lamuda}, where both the overall cooperation level and the system's average reputation increase significantly as $r_0$ increases.

Under strong social dilemma conditions ($r_0 = 2$), when $\beta$ and $\lambda$ are small, both the cooperation frequency and the average reputation decline rapidly, and the system eventually settles into a full-defection state characterized by uniformly low reputation levels. In contrast, when $\beta$ and $\lambda$ are relatively large, such as in Figs.~\ref{fig:lamuda}(\subref{fig:lamuda_2_fc}) and~\ref{fig:lamuda}(\subref{fig:lamuda_2_R}) where $\beta=2$ and $\lambda=2$, cooperators gradually cluster based on their reputation advantages, forming stable cooperative blocks. These blocks, benefiting from the multiplication factor enhanced by reputation, achieve higher payoffs and expand cooperative regions through imitation, thus significantly elevating the system's overall cooperation level.

Under moderate dilemma conditions ($r_0 \approx 3$), cooperation can still be partially sustained even when $\beta$ and $\lambda$ take intermediate values. Moreover, larger values of $\beta$ and $\lambda$ further accelerate the accumulation of reputation and the spread of cooperation, allowing the system to rapidly converge to a steady state with high cooperation.

Under weak social dilemma conditions ($r_0 > 5$), the large value of $r_0$ enhances the group payoff derived from cooperation, which in turn makes cooperation more stable. As a result, high levels of cooperation and reputation can be maintained in all combinations. Furthermore, the comparison further demonstrates that positive payoff amplification ($\beta>0$) and high-reputation imitation preference ($\lambda>0$) effectively lower the social dilemma threshold required for cooperation. The reduction allows cooperation to emerge even under harsher payoff conditions (i.e., lower $r_0$) and significantly accelerates the system’s convergence to a steady state characterized by high cooperation and high reputation. By contrast, negative payoff amplification ($\beta<0$) strengthens the intensity of the social dilemma, which means that a larger $r_0$ is required to sustain global cooperation within the system.

Thus, the synergistic effect of $\beta$ and $\lambda$ is crucial under strong dilemma conditions. When both parameters take small values, the system tends to remain in a state of full defection equilibrium, and cooperation cannot emerge spontaneously. In particular, a large $\beta$ enables individuals with high reputation to gain significant payoff advantages, while a large $\lambda$ makes individuals more inclined to imitate these high-reputation neighbors. These two effects, when combined, can effectively break the equilibrium of defection. Instead, under weak dilemma conditions, the reputation mechanism mainly functions to consolidate cooperation and enhance payoffs. In sum, the proposed assimilated reputation mechanism exhibits generality and robustness at different levels of social dilemma.

\subsection{Spatial evolution of cooperation with reputation}
\begin{figure}
    \centering
    \includegraphics[width=1\linewidth]{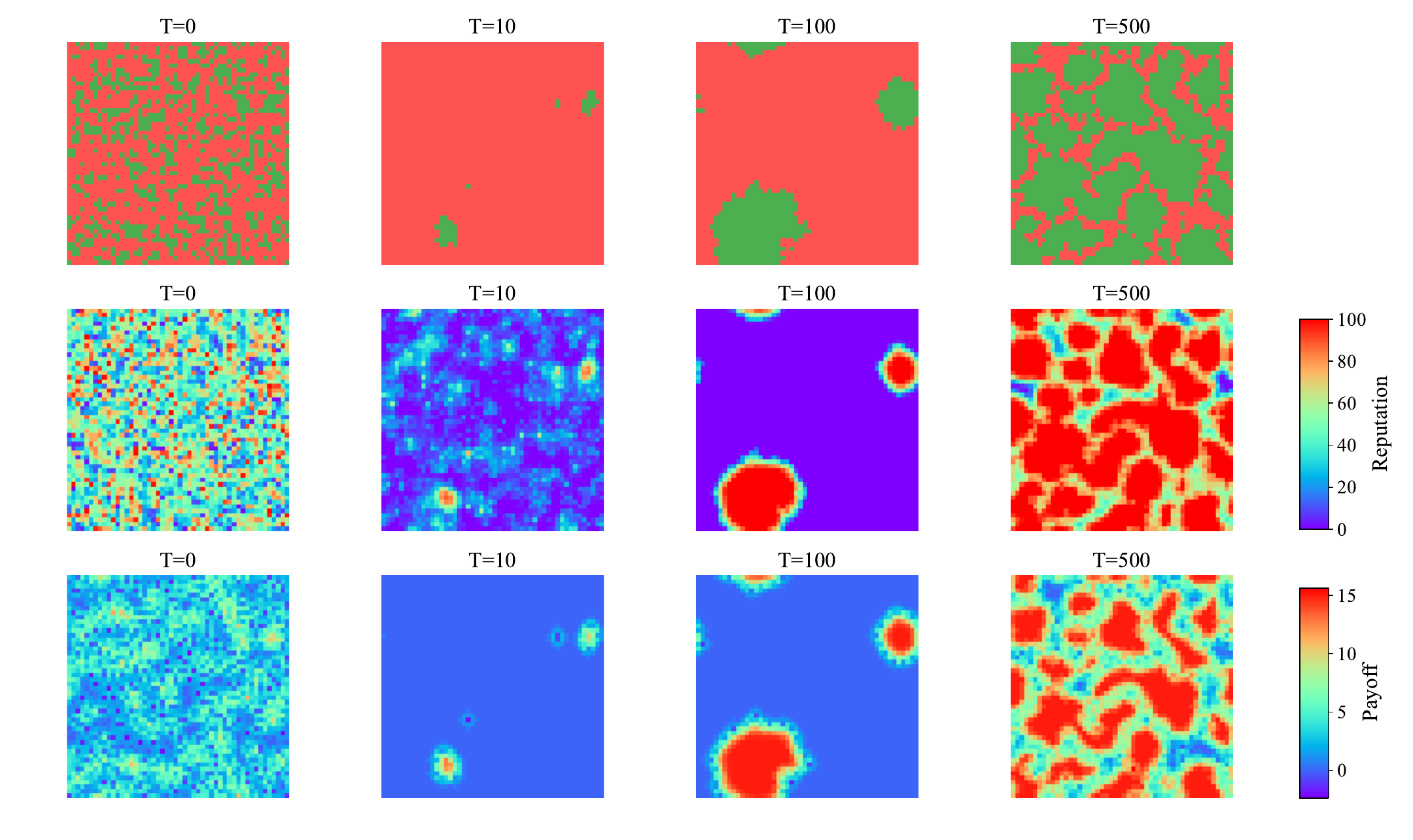}
    \caption{\textbf{Spatial evolution of strategy, reputation, and payoff distributions.} The top row shows the strategy distribution at four representative stages. Here, red (green) represents defector (cooperator) players. The middle and bottom rows show the related reputation and payoff values.
    These snapshots demonstrate a strong interdependence between cooperation, reputation, and payoff values. The parameters are $r_0=2$, $\lambda=2$, $\beta=2$, $\alpha=0.5$, and $\delta=5$.
}
    \label{fig:snapshots}
\end{figure}

However, the average value of reputation and the level of cooperation are insufficient to reveal the strong correlation between these values. To explore how cooperation can emerge from a random initial state and form stable cooperative clusters under strong social dilemma conditions through the assimilated reputation mechanism, representative spatial snapshots are shown to visually depict the distributions of individual strategies, reputations, and payoffs at different times, including $T=0, 10, 100, 500$. They are shown in Fig.~\ref{fig:snapshots}.  

The baseline synergy factor is set to $r_0=2$, while parameters for strong payoff amplification and reputation sensitivity ($\beta=2$ and $\lambda=2$) are selected to explicitly highlight the driving role of reputation under dilemma conditions. At the initial time point ($T=0$), the distribution of individual strategies, reputations, and payoffs is entirely random. By $T=10$, the high intensity of the social dilemma causes defectors to expand rapidly by leveraging their payoff advantages over the short term, which in turn leads to a widespread collapse of cooperation within the system. Consequently, the overall payoff and reputation of the system decrease significantly, with only a small number of regions remaining capable of sustaining high reputation and cooperative behavior.

At $T=100$, a small number of cooperators survive by forming compact clusters that gradually expand along their boundaries. Individuals within these clusters achieve higher payoffs due to improved reputation, thereby establishing spatial survival advantages. As evolution proceeds to $T=500$, cooperative clusters continue to expand along boundaries and merge with neighboring clusters, eventually forming dominant regions characterized by high cooperation, high reputation, and high payoff values. At this stage, the system attains a steady state in which both the cooperation level and the average reputation are maintained at high values.

The positive feedback between reputation and cooperation enables local cooperative clusters to accumulate payoff advantages and resist the invasion of defectors. Moreover, payoff differences drive strategy updates, allowing cooperative clusters to continue their spatial expansion. These spatial dynamics, consistent with the results presented in Fig.~\ref{fig:lamuda}, provide additional evidence that the assimilated reputation mechanism can break the trap of full defection and drive the system towards a steady state with high cooperation.

\subsection{Effects of reputation perturbation factor on cooperation stability}
The effectiveness of the reputation mechanism relies on the extent of the reputation perturbation factor, which governs the responsiveness of the reputation to individual behaviors. When this feedback is too weak, the payoff differences between cooperation and defection are insufficient to alter individual behavioral choices. Conversely, an appropriately strong reputation perturbation factor can serve as a key factor in breaking the equilibrium of full defection.

\begin{figure}
    \centering
    \begin{subfigure}{0.45\linewidth}  
        \centering
        \includegraphics[width=\linewidth]{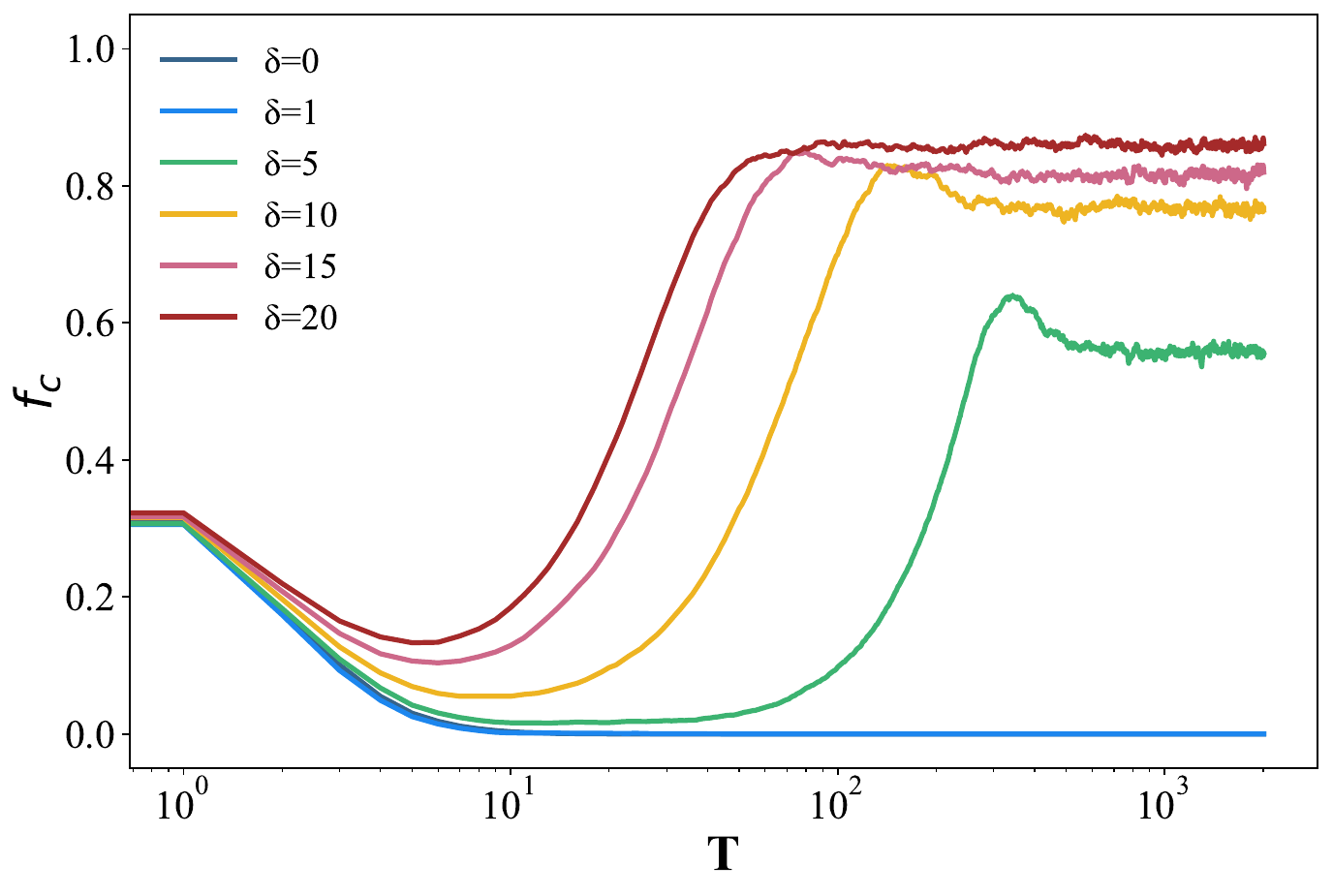}  
        \caption{$f_c$}
        \label{delta.a}
    \end{subfigure}
    \hspace{0.03\linewidth}  
    \begin{subfigure}{0.45\linewidth}  
        \centering
        \includegraphics[width=\linewidth]{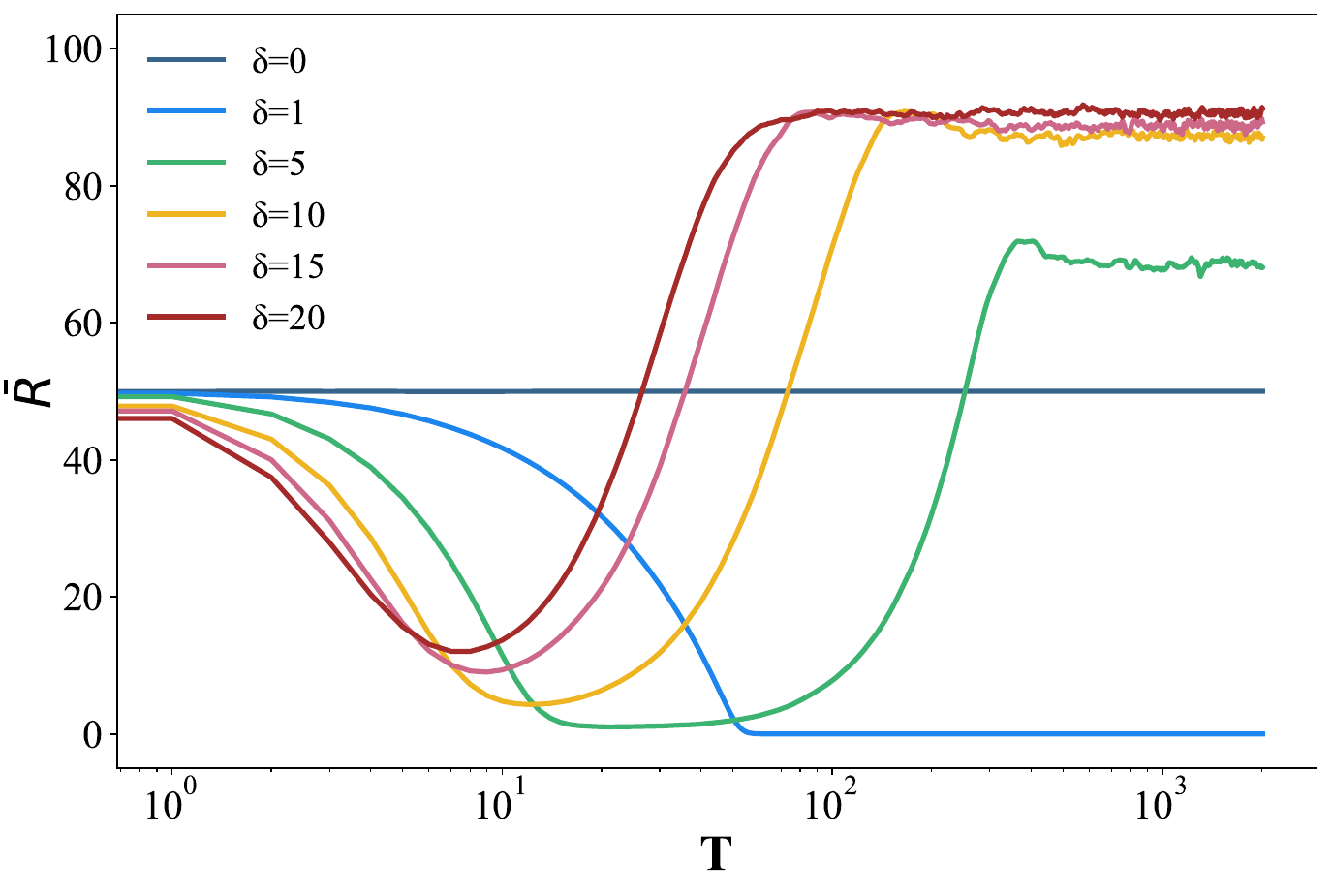}
        \caption{$\bar{R}$}
        \label{delta.b}
    \end{subfigure}
    \caption{\textbf{Evolutionary trajectories of the cooperation fraction $f_c$ and average reputation $\bar{R}$ under different levels of reputation perturbation factor $\delta$.} Different colored curves correspond to different values of $\delta$, as indicated in the legend. Other parameters are fixed as $r_0=2$, $\lambda=2$, $\beta=2$, and $\alpha=0.5$. Panel~(a) illustrates the well-known ``first down, later up'' dynamics of cooperation level, which is a trademark of enhanced network reciprocity~\cite{perc_pre08b,szolnoki_epjb09}. Panel~(b) demonstrates that the average reputation evolves with the same dynamics, which supports our findings shown in Fig.~\ref{fig:snapshots}.}
    \label{fig:delta}
\end{figure}

To examine the effect, the impact of the reputation perturbation factor $\delta$ on the evolution of cooperation is further analyzed, shown in Fig.~\ref{fig:delta}. The results reveal that when $\delta=0$, the absence of the reputation perturbation factor prevents effective differentiation between cooperators and defectors, causing the fraction of cooperators to decline rapidly in the early stages of evolution and driving the system toward a full-defection state. Meanwhile, the average reputation remains close to its initial medium level, exhibiting minimal fluctuations. This finding suggests that without adequate reputation adjustment, reputation becomes decoupled from individual behavior, hindering cooperators from gaining cumulative advantages, and leading the system into a stagnant state characterized by complete defection and moderate reputation. 

Instead, when $\delta>1$, the cooperation fraction initially declines because individuals tend to exploit gains in the short term through defection. However, as $\delta$ increases, both the magnitude and the duration of the early decline are substantially reduced. Subsequently, the cooperation fraction gradually recovers and converges to a stable level that increases with larger $\delta$. The evolution of the average reputation closely mirrors that of the cooperation fraction, initially decreasing but subsequently recovering under the influence of the reputation perturbation factor and ultimately stabilizing at a high level. Notably, cooperation and reputation reach high values in the steady state more rapidly under large $\delta$ conditions.

With a large value of $\delta$, the reputations of defectors decline significantly due to their short-sighted choices, which in turn reduces their payoff values and their attractiveness as targets for imitation in subsequent rounds. Conversely, cooperators accumulate higher reputations through active contributions, gaining additional payoff and becoming more likely targets for imitation. The positive feedback gradually restores cooperation and eventually establishes its dominance within the system. These results demonstrate that only when the reputation perturbation factor is sufficiently strong can the mechanism effectively promote cooperation and suppress defection, which in turn sustains cooperation even under strong social dilemmas.

\subsection{Synergistic impacts of key reputation parameters}
\begin{figure}
    \centering
    \begin{subfigure}{0.285\linewidth}  
        \centering
        \includegraphics[width=\linewidth]{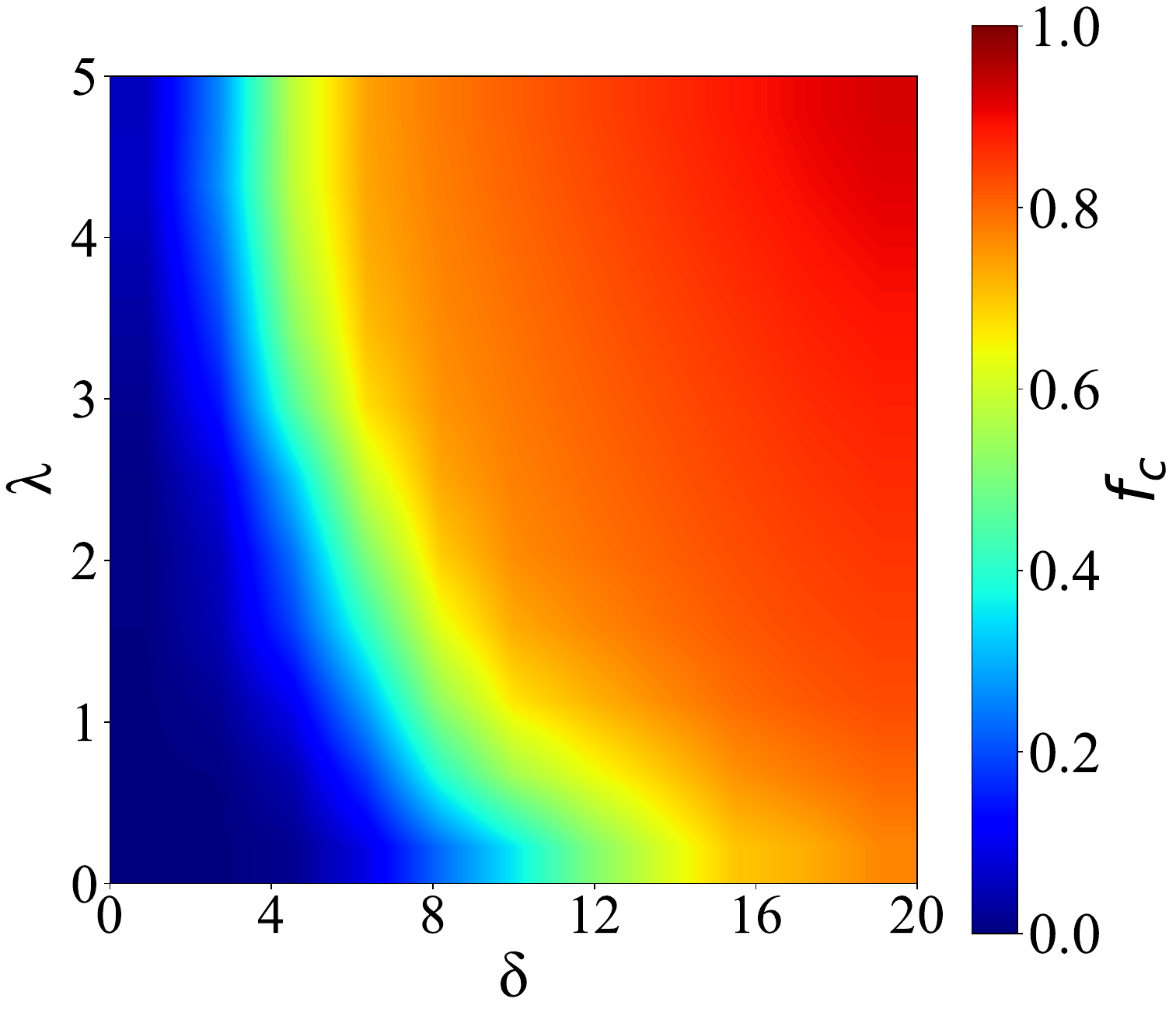}  
        \caption{$f_c$ under $(\lambda,\delta)$}
        \label{subfig:lamuda_deta_cooperation}
    \end{subfigure}
    \hspace{0.03\linewidth}  
    \begin{subfigure}{0.30\linewidth}  
        \centering
        \includegraphics[width=\linewidth]{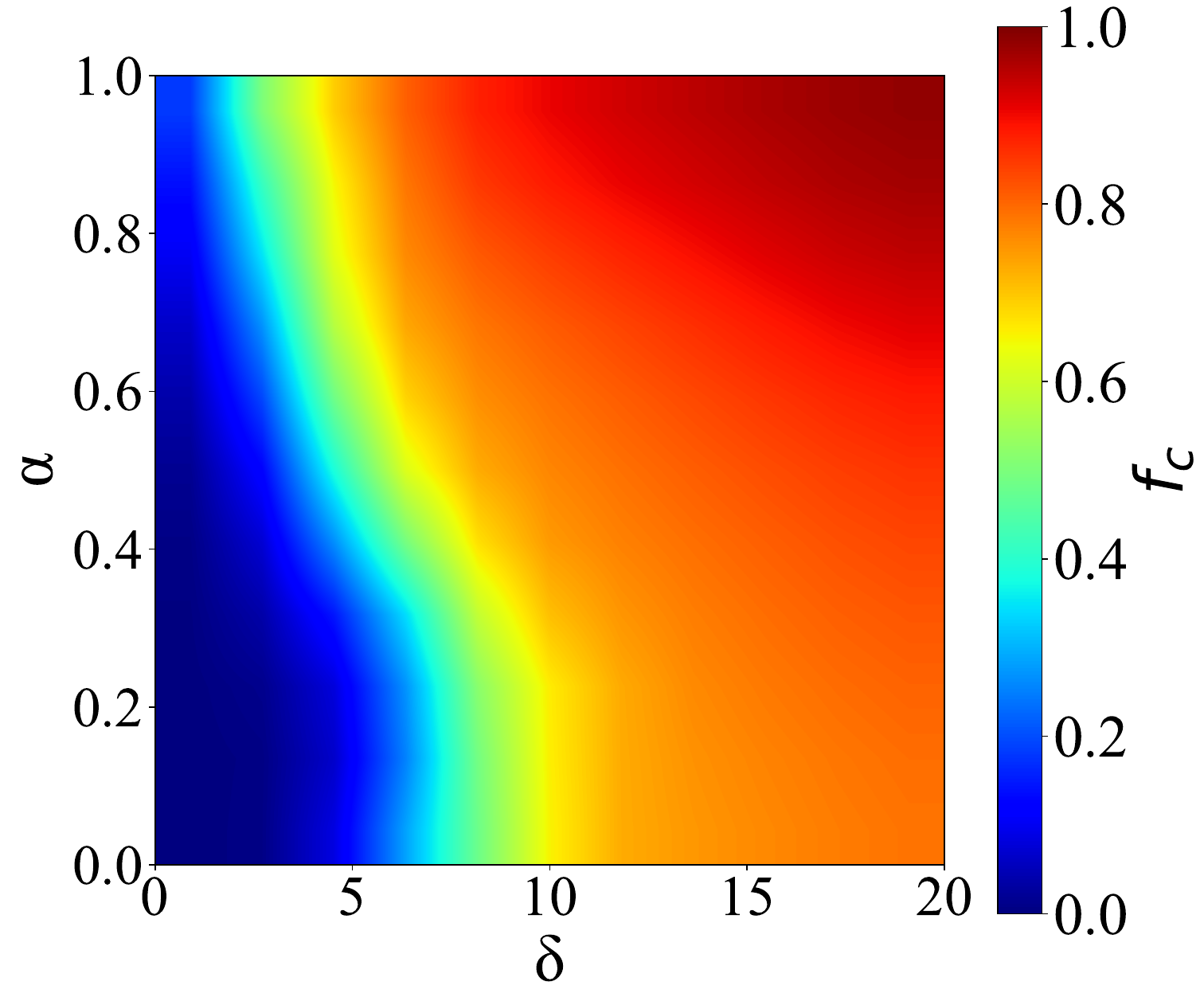}
        \caption{$f_c$ under $(\alpha,\delta)$}
        \label{subfig:alpha_delta_fc}
    \end{subfigure}
    \hspace{0.03\linewidth}  
    \begin{subfigure}{0.30\linewidth}  
        \centering
        \includegraphics[width=\linewidth]{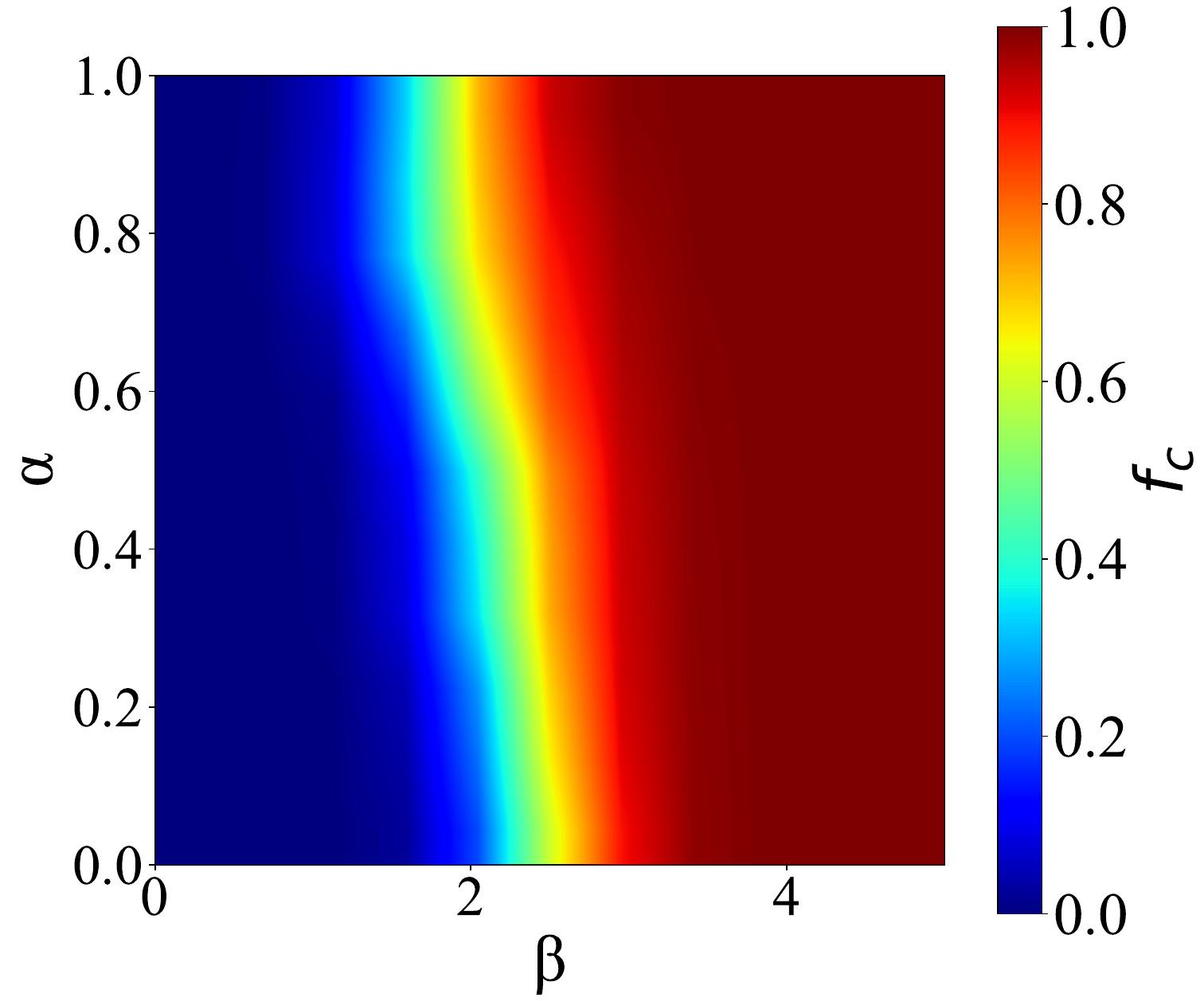}
        \caption{$f_c$ under $(\beta,\alpha)$}
        \label{subfig:alpha_beta_fc}
    \end{subfigure}\\  
    \begin{subfigure}{0.285\linewidth}  
        \centering
        \includegraphics[width=\linewidth]{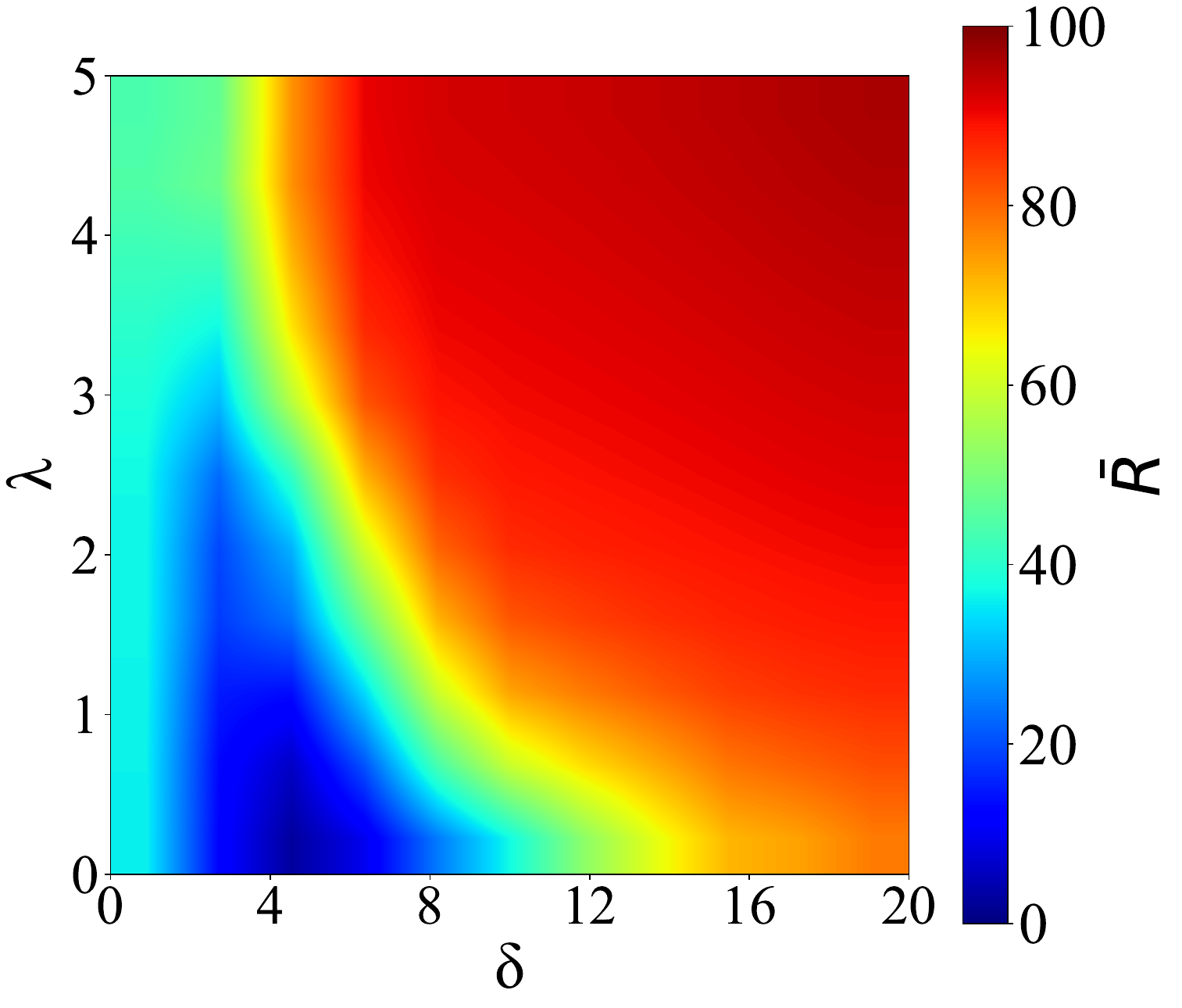}  
        \caption{$\bar{R}$ under $(\lambda,\delta)$}
        \label{subfig:lamuda_deta_reputation}
    \end{subfigure}
    \hspace{0.03\linewidth}  
    \begin{subfigure}{0.30\linewidth}  
        \centering
        \includegraphics[width=\linewidth]{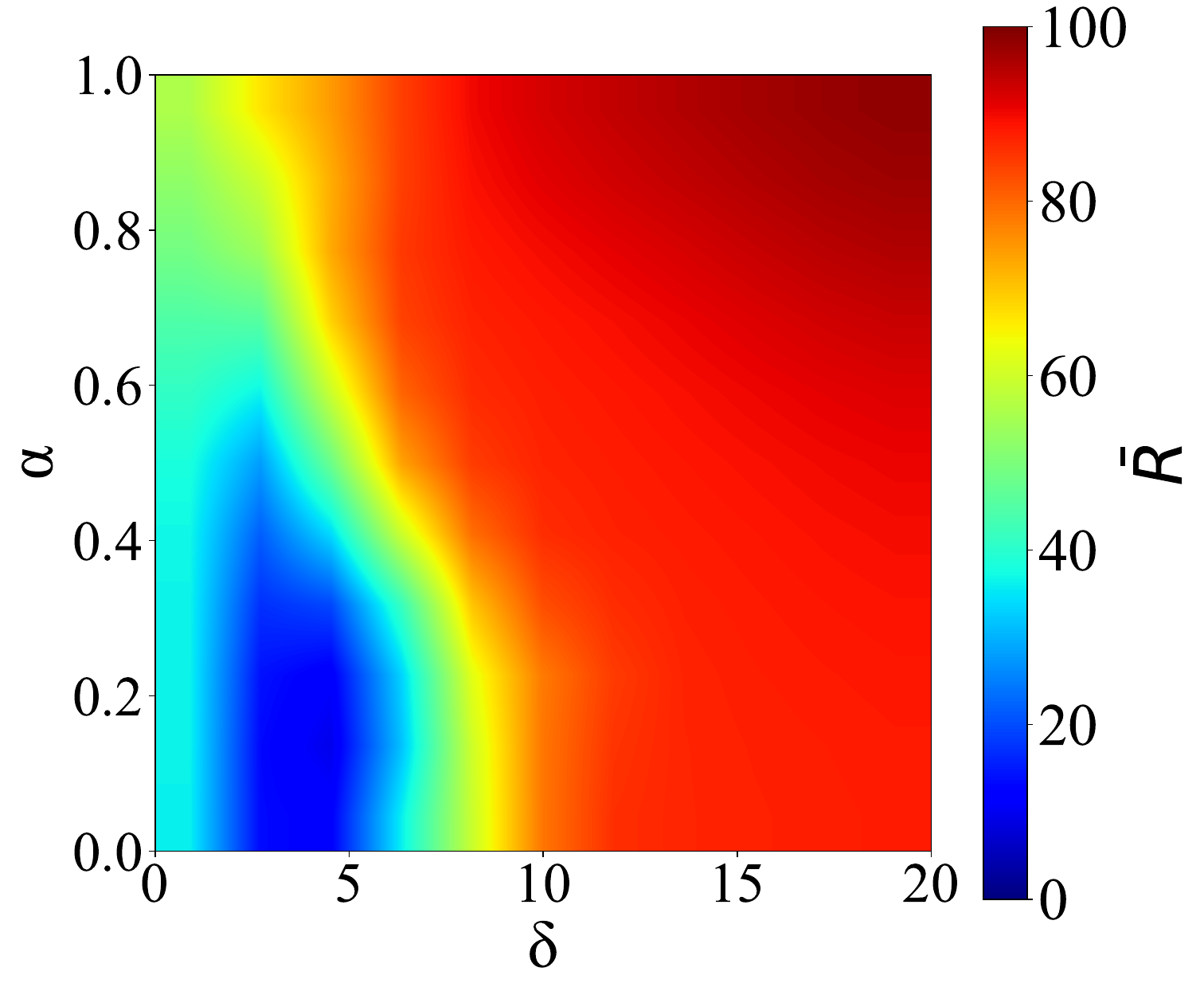}
        \caption{$\bar{R}$ under $(\alpha,\delta)$}
        \label{subfig:alpha_delta_R}
    \end{subfigure}
    \hspace{0.03\linewidth}  
    \begin{subfigure}{0.30\linewidth}  
        \centering
        \includegraphics[width=\linewidth]{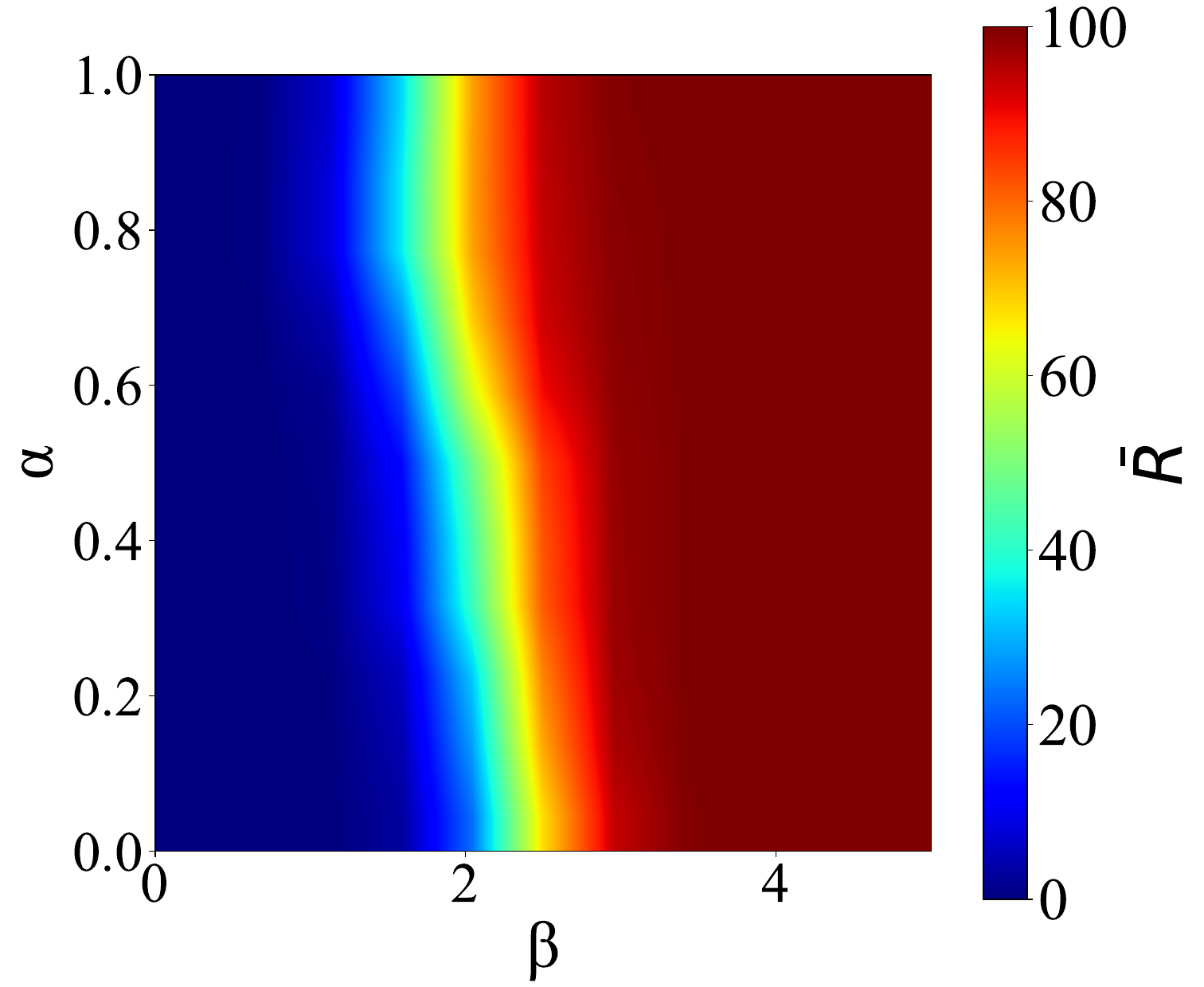}
        \caption{$\bar{R}$ under $(\beta,\alpha)$}
        \label{subfig:alpha_beta_R}
    \end{subfigure}
    \caption{\textbf{Color coded values of stationary cooperation level $f_c$ and average reputation $\bar{R}$ under different parameter combinations.} 
Figs.~\ref{fig:heatmaps_combined}(\subref{subfig:lamuda_deta_cooperation}) and \ref{fig:heatmaps_combined}(\subref{subfig:lamuda_deta_reputation}) show the joint effects of $\lambda$ and $\delta$, 
Figs.~\ref{fig:heatmaps_combined}(\subref{subfig:alpha_delta_fc}) and \ref{fig:heatmaps_combined}(\subref{subfig:alpha_delta_R}) show the joint effects of $\alpha$ and $\delta$, 
and Figs.~\ref{fig:heatmaps_combined}(\subref{subfig:alpha_beta_fc}) and \ref{fig:heatmaps_combined}(\subref{subfig:alpha_beta_R}) show the joint effects of $\beta$ and $\alpha$. Except for the parameters being varied, all other parameters are fixed as $r_0=2$, $\lambda=2$, $\beta=2$, $\alpha=0.5$, and $\delta=5$. }
    \label{fig:heatmaps_combined}
\end{figure}

To further elucidate how the core parameters associated with reputation jointly influence cooperative dynamics, we conduct a series of simulations exploring the pairwise interactions among reputation sensitivity ($\lambda$), reputation perturbation factor ($\delta$), reputation assimilation coefficient ($\alpha$), and payoff amplification ($\beta$). The stable distributions of the cooperation level $f_c$ and the average reputation $\bar{R}$ under different combinations of the parameters are presented in Fig.~\ref{fig:heatmaps_combined}. These results provide a comprehensive view of how the feedback mechanisms rooted in reputation assimilation simultaneously govern the emergence, spread, and stabilization of cooperation.

Figs.~\ref{fig:heatmaps_combined}(\subref{subfig:lamuda_deta_cooperation}) and \ref{fig:heatmaps_combined}(\subref{subfig:lamuda_deta_reputation}) illustrate the coupled effects of reputation sensitivity $\lambda$ and reputation perturbation factor $\delta$. We find that increasing $\lambda$ significantly enhances the tendency of individuals to imitate a highly reputable neighbor, accelerating the spread of cooperative strategies within the population. When $\lambda$ is relatively large, the system can quickly attain a steady state characterized by high cooperation and high reputation, even if $\delta$ is relatively low. Conversely, if $\lambda$ is small or approaches 0, the effect of the magnitude of the reputation perturbation factor is weakened, and the system remains in a low cooperation and low reputation state for an extended period under low $\delta$. When $\delta$ is close to zero, the system lacks effective feedback, making it difficult for cooperation to persist. As a result, the cooperation frequency rapidly drops to a very low level, while the average reputation remains around its initial moderate value. This outcome is consistent with the case of $\delta=0$ shown in Fig.~\ref{fig:delta}.

The interplay between the reputation assimilation coefficient $\alpha$ and the reputation perturbation factor $\delta$ is shown in Figs.~\ref{fig:heatmaps_combined}(\subref{subfig:alpha_delta_fc}) and \ref{fig:heatmaps_combined}(\subref{subfig:alpha_delta_R}). The figures show that when $\delta$ is close to zero, the lack of an effective reputation perturbation factor prevents the maintenance of cooperation. Consequently, the fraction of cooperators declines rapidly to a very low level, while the average reputation stays around its initial moderate value. The phenomenon is consistent with the results for $\delta=0$ presented in Fig.~\ref{fig:delta}. As $\delta$ increases, the cooperation level rises gradually, indicating a transition from a low to a high cooperative state. Notably, the magnitude of the reputation assimilation coefficient $\alpha$ significantly shifts the critical point of the transition. When $\alpha$ is large, historical reputation exerts a stronger influence on individuals' current reputation, making reputation accumulation more stable and less susceptible to short-term fluctuations. The stability enables cooperation to be initiated and maintained even when $\delta$ is relatively low. In contrast, when $\alpha$ is small, the individual's reputation is easily disturbed by immediate behaviors. In such cases, a stronger reputation perturbation factor is required to overcome the defection trap and guide the system toward a stable cooperative state.

Finally, Figs.~\ref{fig:heatmaps_combined}(\subref{subfig:alpha_beta_fc}) and \ref{fig:heatmaps_combined}(\subref{subfig:alpha_beta_R}) demonstrate the combined effects of payoff amplification $\beta$ and the reputation assimilation coefficient $\alpha$. When $\beta=0$, the reputation-induced amplification effect in the public goods game disappears completely. Cooperation does not receive any additional incentive in this case, which causes the system to rapidly degenerate into a state where all individuals are defectors and the reputation is 0. As $\beta$ gradually increases, the fraction of cooperators undergoes a pronounced transition, rapidly rising from low to high levels. The average reputation increases correspondingly and approaches 100 when $\beta$ reaches approximately 5. Noticeable differences in the growth rates of cooperation and reputation are observed at different values of $\alpha$. When $\alpha$ is large, the historical reputation of individuals persists over longer time scales, which allows both the frequency of cooperation and the average reputation to reach a higher level under the same $\beta$. In contrast, when $\alpha$ is small, although the system still transitions from low to high levels of cooperation and reputation as $\beta$ increases, the growth process is slower and more susceptible to short-term fluctuations.

These results reveal that the four key parameters jointly govern the emergence and stabilization of cooperation. The reputation perturbation factor $\delta$ amplifies the reputation gap between cooperators and defectors, breaking the Nash equilibrium of full defection and initiating the transition towards cooperation. The reputation sensitivity $\lambda$ enforces individuals to imitate high-reputation neighbors, accelerating the diffusion of cooperative strategies throughout the population. The reputation assimilation coefficient $\alpha$ preserves historical evaluations and mitigates short-term fluctuations, ensuring long-term stability of cooperative states. Meanwhile, the payoff amplification factor $\beta$ links reputation to higher group payoffs, providing additional incentives that reinforce positive feedback between cooperation and reputation. Through the synergistic effects of these parameters, the system ultimately evolves toward a stable configuration characterized by persistently high cooperation and reputation.

\subsection{Robustness of the assimilated reputation mechanism across network topologies}

\begin{figure}
    \centering
    \begin{subfigure}[b]{0.45\textwidth}
        \centering
        \includegraphics[width=\linewidth]{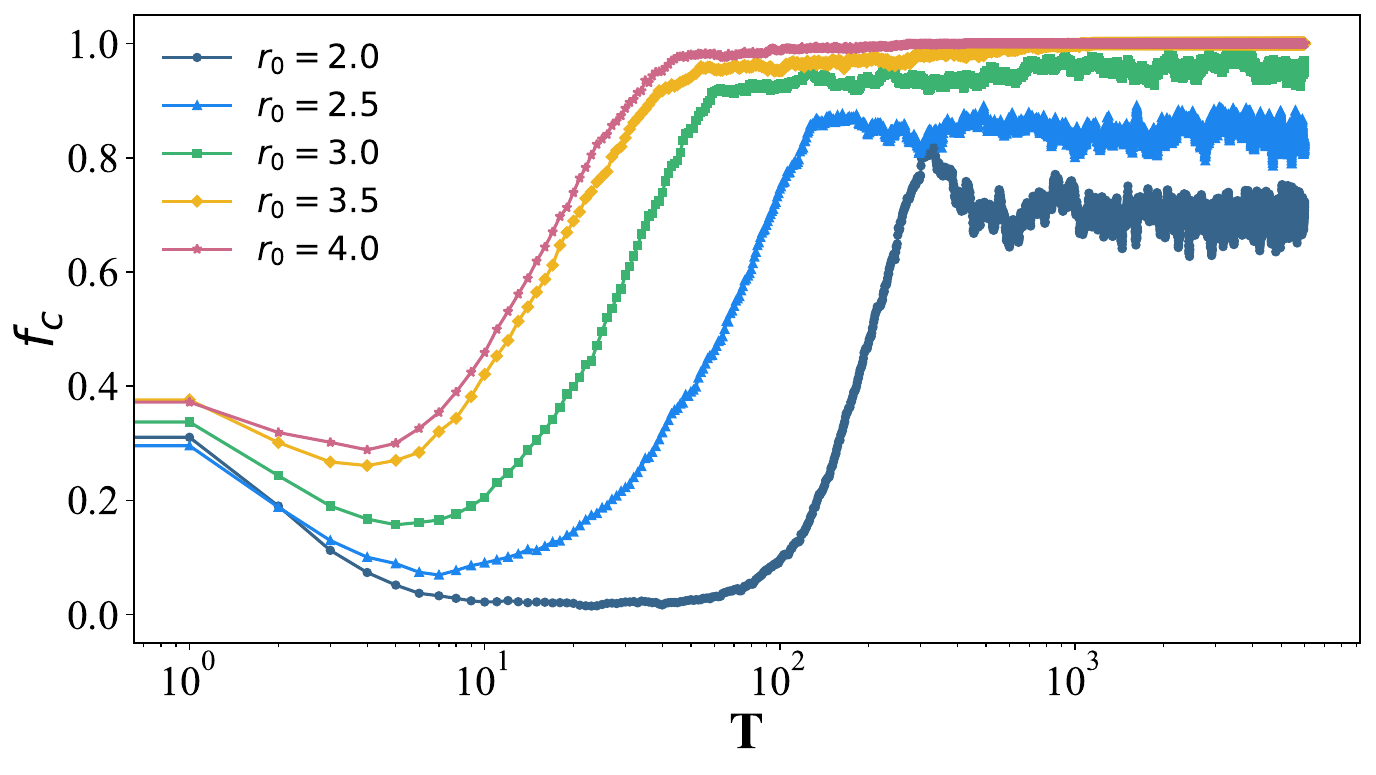}
        \caption{$SL$}
        \label{fig:SL_fc}
    \end{subfigure}
    \hfill
    \begin{subfigure}[b]{0.45\textwidth}
        \centering
        \includegraphics[width=\linewidth]{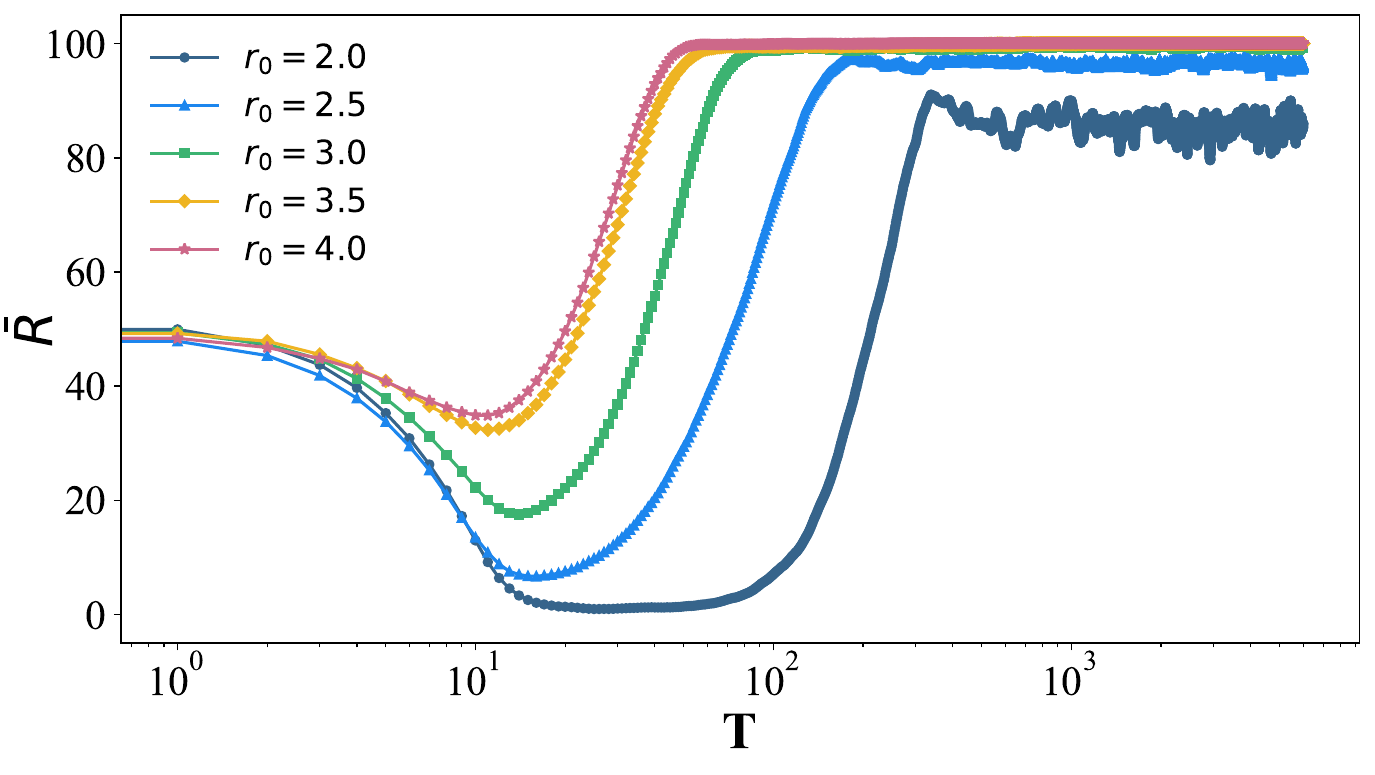} 
        \caption{$SL$}
        \label{fig:SL_R}
    \end{subfigure}
    
    \vspace{1em} 

    \begin{subfigure}[b]{0.45\textwidth}
        \centering
        \includegraphics[width=\linewidth]{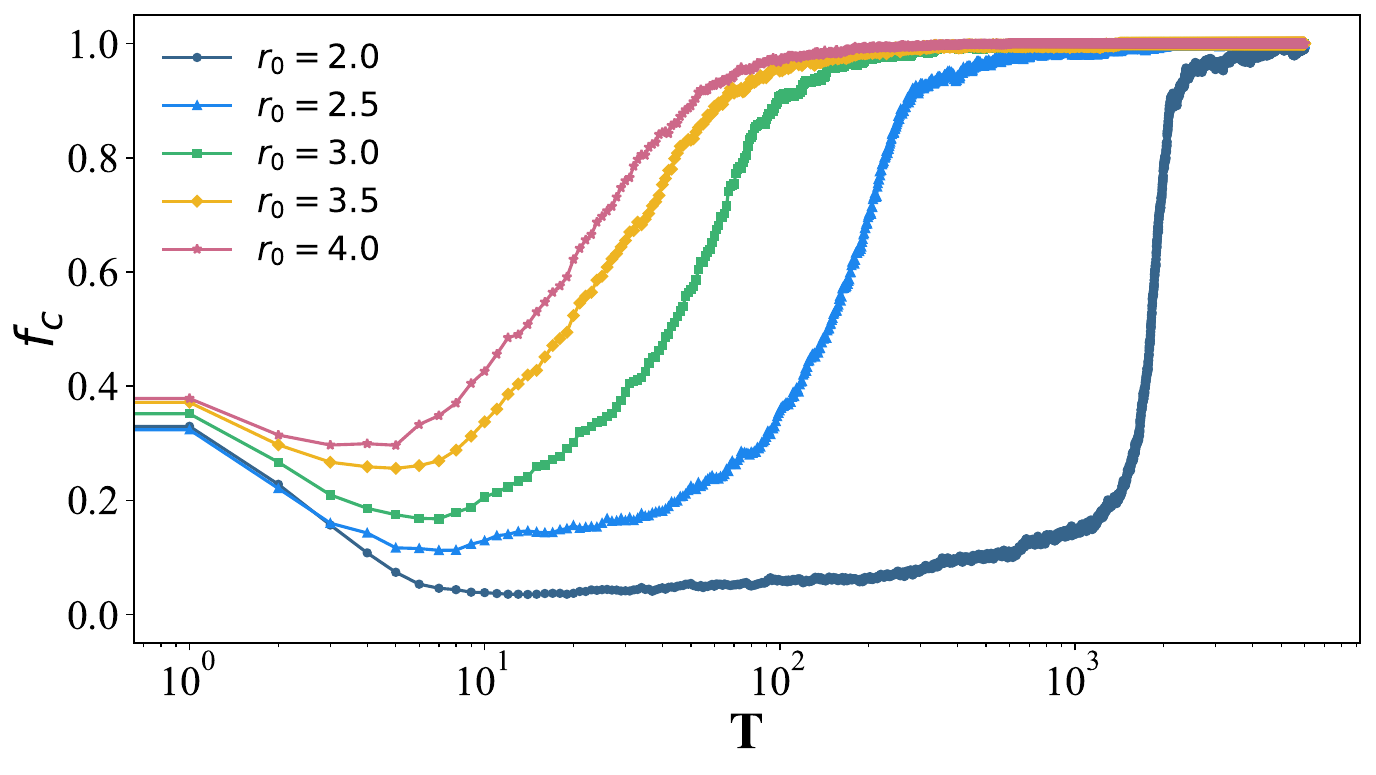}
        \caption{$WS$}
        \label{fig:WS_fc}
    \end{subfigure}
    \hfill
    \begin{subfigure}[b]{0.45\textwidth}
        \centering
        \includegraphics[width=\linewidth]{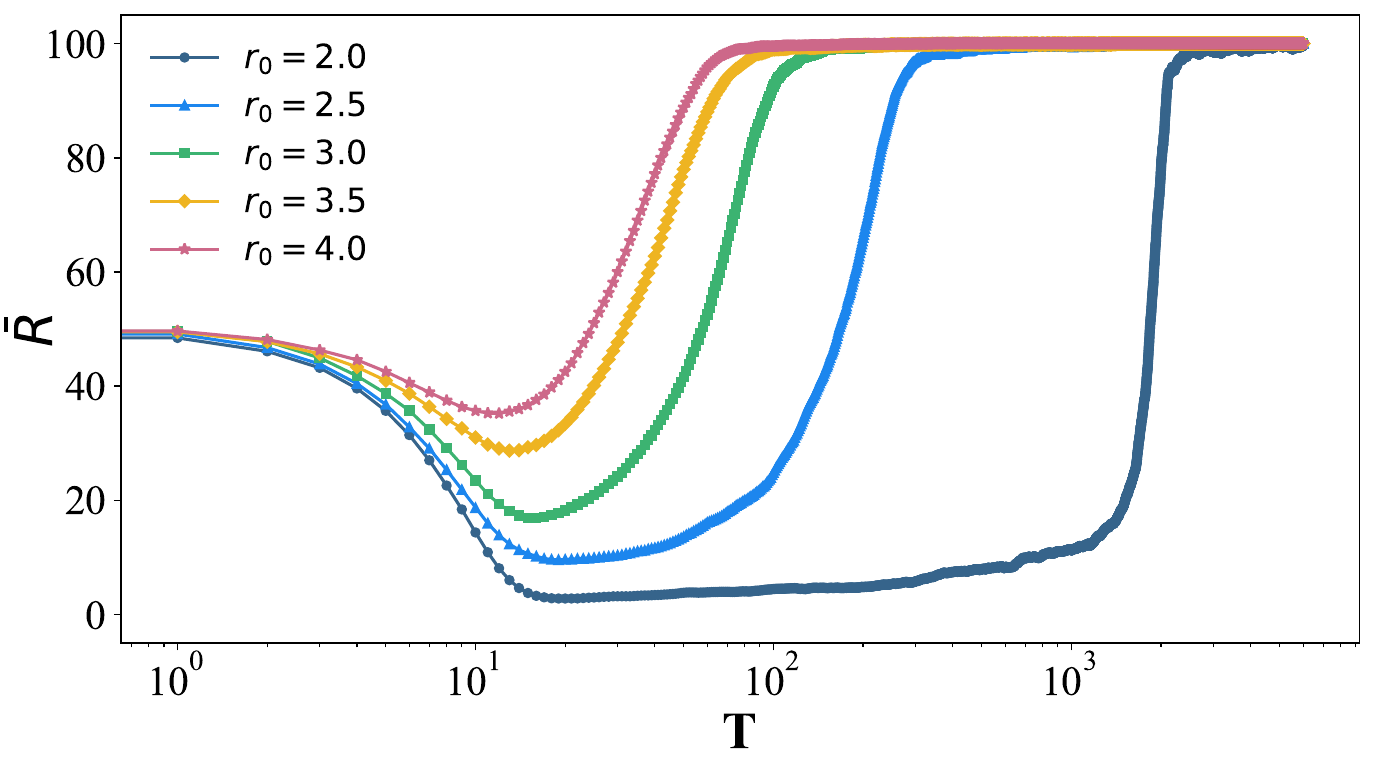} 
        \caption{$WS$}
        \label{fig:WS_R}
    \end{subfigure}
    \caption{\textbf{Temporal evolution of the $f_c$ and $\bar{R}$ on square lattice and WS small-world networks under different $ r_0$.} Figs.~\ref{fig:ws}(\subref{fig:SL_fc}) and~\ref{fig:ws}(\subref{fig:SL_R}) correspond to the square lattice, while Figs.~\ref{fig:ws}(\subref{fig:WS_fc}) and~\ref{fig:ws}(\subref{fig:WS_R}) present results on the WS small-world network. 
    Figs.~\ref{fig:ws}(\subref{fig:SL_fc}) and~\ref{fig:ws}(\subref{fig:WS_fc}) depict the time evolution of the cooperation fraction $f_c$, and Figs.~\ref{fig:ws}(\subref{fig:SL_R}) and~\ref{fig:ws}(\subref{fig:WS_R}) show the corresponding average reputation $\bar{R}$. Different colored curves represent varying baseline multiplication factors of $r_0$: dark blue circles for $r_0=2.0$, blue triangles for $r_0=2.5$, green squares for $r_0=3.0$, yellow diamonds for $r_0=3.5$, and red pentagrams for $r_0=4.0$. }
    \label{fig:ws}
\end{figure}

To further check the robustness of our observations based on the proposed assimilated reputation mechanism, we compare the results obtained on square lattice (SL) and non-regular WS small-world networks (WS). While the square lattice allows local interactions only, the WS topology introduces connections that span distant nodes with a rewiring probability $p=0.1$, capturing the mixed structure of real social systems that combine local clustering with occasional global connectivity.

As shown in Fig.~\ref{fig:ws}, the temporal trajectories reveal that as time progresses, the fraction of cooperators $f_c$ and the average reputation $\bar{R}$ experience an initial decline followed by recovery and stabilization at high levels. This trend is consistent across both network structures, indicating that the proposed assimilated reputation mechanism effectively sustains cooperation over time. Furthermore, increasing the baseline multiplication factor $r_0$ consistently enhances cooperative clustering and accelerates reputation accumulation, resulting in higher steady levels of $f_c$ and $\bar{R}$ on both topologies.

Notably, under weak social dilemmas, the evolutionary trajectories on the WS small-world network and the square lattice almost coincide, with both systems rapidly converging to high levels of cooperation and reputation. Under strong social dilemmas, however, the formation of cooperative clusters on the WS network occurs more slowly than on the lattice, as the additional shortcut links enhance node mixing and hinder the early development of local cooperative domains. Once these clusters are fully established, the WS topology generally achieves a slightly higher stable cooperation level than the lattice. This observation is consistent with the well-known property of small-world networks, which can optimize cooperation by balancing clustering and path length \cite{masuda2003spatial}. These results further demonstrate that, in the presence of the assimilated reputation mechanism, the WS network can achieve cooperation levels comparable to or even exceeding those of the lattice network, with differences mainly arising during the initial phase of cooperative formation.

Summing up, these results confirm that the assimilated reputation mechanism maintains its effectiveness in fostering cooperation on distinct network topologies. Its ability to generate and sustain high levels of cooperation and reputation remains intact in both regular and random structures, demonstrating strong structural robustness and universality in complex social environments.

\section{Conclusions and outlook} \label{Conclusions}
In this paper, we have studied the consequences of an extended reputation that can be interpreted as a sort of reputation assimilation. This extension describes a general observation that the reputation of an individual depends not only on their past behavior, but it is also influenced by the actions of their close neighborhood. In this way, there is a sophisticated interaction between an individual and group activities. As a consequence, the extended reputation simultaneously regulates the enhancement factor of collective contributions and the willingness of players to cooperate, while the group evolution is jointly shaped by local interactions and dynamic updating processes. 
Our model's innovation lies in its success in breaking through the limitation that reputation only serves to describe individual performance in traditional models. Our model also integrates reputation into both strategy imitation and payoff distribution simultaneously, creating a positive feedback that facilitates the emergence and long-term maintenance of cooperation. Through numerical simulations, the effects of reputation assimilation coefficient, reputation perturbation factor, reputation sensitivity, and payoff amplification on cooperative behavior were systematically investigated, and the resulting synergistic interactions among these mechanisms were revealed. Monte Carlo simulations demonstrated that the assimilated reputation mechanism significantly promotes cooperation at different intensity levels of the social dilemma. The applied reputation assimilation coefficient improves stability by preserving historical evaluations, while the reputation perturbation factor prevents premature convergence and maintains diversity. Moreover, the synergy between payoff amplification and reputation sensitivity plays a decisive role under strong dilemmas, enabling the system to overcome the trap of full defection and achieve steady high-cooperation states. We have also conducted simulations on the WS small-world network that confirmed that the assimilated reputation mechanism exhibits consistent cooperative enhancement effects, thereby demonstrating its robustness across different network topologies.

Our present results open new research directions for future studies. Although the present study validates the robustness of the assimilated reputation mechanism in different networks, further investigations could explore more dynamic and heterogeneous environments. For example, extending the framework to temporal or multilayer networks would allow the examination of how evolving connections and interactions across layers influence cooperation dynamics. Additionally, incorporating stochastic perturbations into reputation updates may help model environmental noise and uncertainty more realistically, thereby enhancing the robustness and applicability of the proposed mechanism.

\section*{Acknowledgments}
S. He, Q. Li, and M. Feng are supported by grant No. 62206230 funded by the National Natural Science Foundation of China (NSFC), and grant No. CSTB2023NSCQ-MSX0064 funded by the Natural Science Foundation of Chongqing. A. Szolnoki is supported by the National Research, Development and Innovation Office (NKFIH) under Grant No.~K142948.

\printcredits



\end{document}